\documentclass[reprint, amsmath,amssymb,aps]{revtex4-2}
\usepackage{blindtext}
\usepackage{graphicx}
\usepackage{dcolumn}
\usepackage{hyperref}
\usepackage{mathtools}
\usepackage{xcolor}
\usepackage{wrapfig}
\input{insbox}

\hypersetup{
    colorlinks,
    linkcolor={blue},
    citecolor={blue},
    urlcolor= {blue}
}
\DeclareMathOperator{\sgn}{\text{sgn}}

\DeclarePairedDelimiter\abs{\lvert}{\rvert}

\DeclarePairedDelimiter\floor{\lfloor}{\rfloor}
\newcommand\wpfoi{\tilde{\eta}} 
\newcommand\wpfo{\eta} 
\newcommand\wpfi{\tilde{\xi}} 
\newcommand\wpf{\xi} 
\newcommand\wpft{\dot{\xi}} 
\newcommand\wpfit{\dot{\tilde{\xi}}}  
\newcommand\hsub{}
\newcommand\Pf{P}
\newcommand\iPf{P^{-}}
\newcommand\hi{h^{-}}
\newcommand\thi{\tilde{h}^{-}}
\newcommand\ip{p^{-}}
\newcommand\xg{d_x} 
\newcommand\yg{d_y} 
\newcommand\fPf{\tilde P} 
\newcommand\gr{q} 
\newcommand\prs{q}
\newcommand\iprs{q^{-}}
\newcommand\mkz{k_{z}}  
\newcommand\mky{k_{y}}  
\newcommand\mkx{k_{x}}  
\newcommand\mko{k}  
\newcommand\moo{\omega}  
\newcommand\rpx{0}
\newcommand\mrpx{}
\newcommand\lhx[1]{x^{(#1)}} 
\newcommand\lhy[1]{y^{(#1)}} 
\newcommand\lpx[1]{x^{(#1)}_{\omega}} 
\newcommand\lpxomin[1]{x^{(#1)}_{\omega_{min}}} 
\newcommand\lpxomax[1]{x^{(#1)}_{\omega_{max}}} 
\newcommand\lpxoo[1]{x^{(#1)}_{\omega_{0}}} 
\newcommand\cen[1]{\langle{#1}\rangle}
\newcommand{\ft}[1]{\text{FT}{\left[#1\right]} }

\begin{document}
\preprint{APS/123-QED}
\title{Fundamentals of Quantum Fourier Optics}
\author{Mohammad Rezai}
\email{m.rezai@ee.sharif.edu}
\altaffiliation{\textit{Sharif Quantum Center} (SQC), Sharif University of Technology, Tehran, Iran}
\altaffiliation{Electrical Engineering Department, Sharif University of Technology, Tehran, Iran}
\author{Jawad A. Salehi}
\email{jasalehi@sharif.edu}
\altaffiliation{\textit{Sharif Quantum Center} (SQC), Sharif University of Technology, Tehran, Iran}
\altaffiliation{Electrical Engineering Department, Sharif University of Technology, Tehran, Iran}
\date{\today}
\maketitle
\textbf{
\textit{Abstract}--All-quantum signal processing techniques are at the core of the successful advancement of most information-based quantum technologies. This paper develops coherent and comprehensive methodologies and mathematical models to describe Fourier optical signal processing in full quantum terms for any input quantum state of light. We begin this paper by introducing a spatially two-dimensional quantum state of a photon, associated with its wavefront and expressible as a two-dimensional creation operator. Then, by breaking down the Fourier optical processing apparatus into its key components, we strive to acquire the quantum unitary transformation or the input/output quantum relation of the two-dimensional creation operators. Subsequently, we take advantage of the above results to develop and obtain the quantum analogous of a few essential Fourier optical apparatus, such as quantum convolution via a 4f-processing system and a quantum 4f-processing system with periodic pupils.
Moreover, due to the importance and widespread use of optical pulse shaping in various optical communications and optical sciences fields, we also present an analogous system in full quantum terms, namely quantum pulse shaping with an 8f-processing system. Finally, we apply our results to two extreme examples of the quantum state of light. One is based on a coherent (Glauber) state and the other on a single-photon number (Fock) state for each of the above optical systems. We believe the schemes and mathematical models developed in this paper can impact many areas of quantum optical signal processing, quantum holography, quantum communications, quantum radars and multiple-input/multiple-output antennas, and many more applications in quantum computations and quantum machine learning algorithms.
}
      \vspace{2em}
\section{Introduction}           
        The extraordinary success of Fourier optics in sciences and engineering throughout the past half-century compel us to return to the basics and grassroots of its ubiquitous and fundamental concepts and strive to dig out its fully quantum analogous~\cite{goodman_2005,poon_2014,papen_blahut_2019}.
        We believe that fully quantum Fourier optic will be considered supreme, like its classical counterpart, in many futuristic and disruptive quantum technologies in information, communications and computations~\cite{knill_n_2001,lukens_o_2017,cariolaro_2015,gisin_rmp_2002}.
        It is from this viewpoint we have researched and written this paper.
        The primary purpose of this paper is to return to the grassroots concepts in Fourier optical signal processing to produce its quantum analogous in component by component basis.
        And subsequently integrate these components to design a complete quantum optical signal processing apparatus tailored to any input quantum lightwave signal.
        \par
        In this paper the information encoding platform is the spatial mode (image) or the wavefront of photons.
        Spatial modes provide a vast quantum information capacity, such as qudits and continuous-mode quantum information for a single photon.
        Photon-wavefront has lately been exploited in photons' orbital angular momentum encoding technique~\cite{mair_n_2001,arnold_pra_2002,fickler_s_2012,fickler_pnas_2016}.
        Moreover, it plays the principal role in quantum imaging~\cite{moreau_nrp_2019,lemos_n_2014,dangelo_lpl_2005,malik_prl_2010,abouraddy_prl_2004,aspden_njp_2013,white_pra_1998,pitman_pra_1996} and quantum holography~\cite{defienne_nphy_2021,abouraddy_oe_2001} and can be used to implement arbitrary and programable unitary transformation~\cite{morizur_osa_2010,lopez_oe_2021}.
        Accordingly, quantum Fourier optics, the core of quantum information processing on photon-wavefront, require special investigation, and it is the main focus of this paper.
        Quantum Fourier optics shows how a spatially two-dimensional quantum state of light (quantum optical signal) evolves through a Fourier optical system.        
        \par
        This paper begins by developing the fundamentals and the building blocks of quantum Fourier optics for a general case of continuous spatial mode.
        Primarily it introduces the quantum state and its photon-wavepacket creation operator in a two-dimensional spatial domain.
        Afterward, it details the quantum model of Fourier optics elements which are free propagation and spatial modulating devices such as lens and diffraction grating. These elements can be combined to create various Fourier optical processors such as the 4f-processing system, bringing unique functionalities such as continuous mode quantum Fourier transformation and quantum convolution.
        It also provides simulation examples for the evolution of the photon-wavepacket through the introduced quantum optical systems.  
         \par
         To demonstrate the supreme capabilities of quantum Fourier optics, as an example, we detail a quantum pulse shaping technique. We show how an 8f-processing system applies the desired phase shifts to each frequency component of a quantum state of light. This technique is used in optical frequency combs quantum information processing~\cite{kues_np_2019}, and quantum code division multiple access (QCDMA) communication systems~\cite{rezai_ieeeit_2021}.
        \section{Generic Representation of the Quantum State of Light as a Function of Creation Operators}
        Fock states provide an appropriate Hilbert space to represent the quantum state of light.
        This paper considers pure quantum light states composed of \textit{identical photons} (some authors call them indistinguishable photons), meaning photons occupying identical mode~$\wpf$.
        Such a quantum state is mathematically representable in the Fock space as follows~\cite{louisell_1990,rezai_ieeeit_2021}
        \begin{equation}
          \begin{split}
            \lvert \psi \rangle &= \sum_{n=0}^{\infty}c_n \lvert n \rangle_{\wpf}=\sum_{n=0}^{\infty} c_n \frac{\hat a_{\wpf}^{\dagger n}}{\sqrt{n!}}\lvert 0 \rangle  =f(\hat a_{\wpf}^ \dagger) \lvert 0 \rangle \, ,
          \end{split}
          \label{eqn:si=f(adag)}
        \end{equation}
        where $\lvert 0 \rangle$ denotes the vacuum state, $\hat a_{\wpf}^\dagger$ is the single-photon creation operator at the generic optical mode~$\wpf$, and $c_n$s fulfill the normalization condition $\sum_{n=0}^{\infty }|c_n|^2=1$ and are the probability amplitudes for the quantum light to be in the photon number states $\lvert n \rangle_{\wpf}$ and associate with the Taylor coefficients of the analytic, infinitely differentiable function $f(\hat a_{\wpf}^ \dagger )=\sum_n c_n \frac{\hat a_{\wpf}^{\dagger n}}{\sqrt{n!}}$.
        It is worth noting that the coherent (Glauber) state~$\lvert \alpha \rangle$ is representable by the exponential function $f(\hat a_{\wpf}^ \dagger)=\exp( -|\alpha|^ 2/2 + \alpha \hat a_{\wpf}^ \dagger)$, where $c_n=\exp( -|\alpha|^ 2/2 ) \frac{\alpha^{n}}{\sqrt{n!}} $, which is equivalent to the displacement operator~\cite{loudon_2000}.
        The number (Fock) state~$\lvert n \rangle_{\wpf}$ is representable by the $n$th order power function, $f(\hat a_{\wpf}^ \dagger)=\frac{1}{\sqrt{n!}}\hat a_{\wpf}^{\dagger n}$, where $c_n=1$ and $c_{n'}=0$ for all $n'\neq n$.
        \par
\begin{figure}[!t]
          \centering
          \includegraphics[width=\columnwidth]{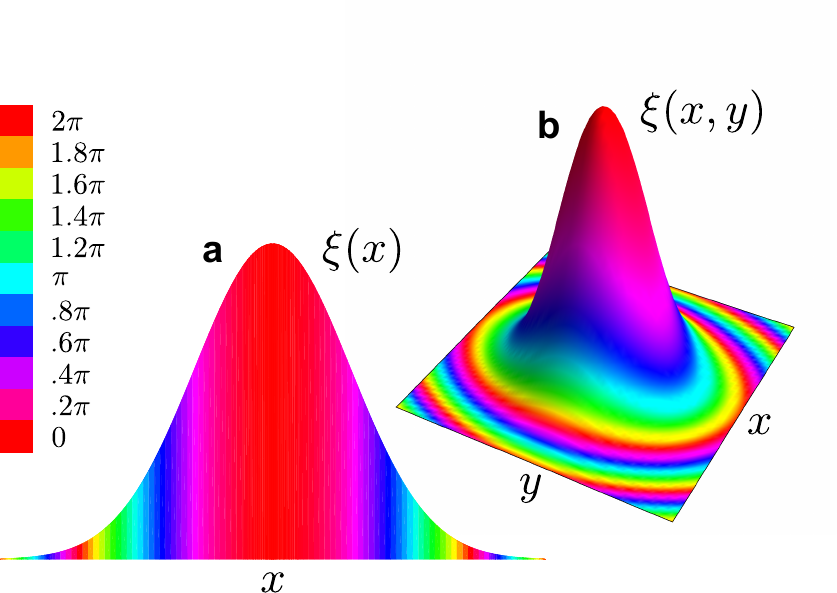}
          \caption{\textbf{Exemplars of photon-wavepackets.}
            \textbf{a} One-dimensional Gaussian photon-wavepacket.
            \textbf{b} Two-dimensional Gaussian photon-wavepacket.           
            The height and the color of the plots indicate the amplitude and the phase (see inset in the left of the figure) of photon-wavepackets, respectively.
          }\label{fig:pwp}
        \end{figure}

        This paper addresses the single-photon wave function as the photon-wavepacket and writes it with the symbol~$\wpf$, which indicates the occupation mode of a single photon.
        As detailed in appendix~\ref{sec:csmco}, the occupation mode’s degree of freedom for a single photon is four, one discrete polarization and three continuous components of the photon wavevector.
        The three wavevector occupation modes can be converted into the photon spectral mode and a two-dimensional spatial mode known as the photon wavefront.
        If the occupation modes of the under-studying photons at any of these four degrees of freedom remain unchanged throughout the system, we drop them from consideration.
        For example, assume the polarization and the wavefront of photons remain intact in the system, and only the spectral mode is under evolution.
        Therefore, the effective photon-wavepacket is reducible to a one-dimensional spectral wavepacket~$\wpf(\omega)$, which can also be Fourier transformed and mapped to a one-dimensional temporal photon wavepacket~\cite{loudon_2000,rezai_ieeeit_2021}.
        A one-dimensional photon wavepacket creation operator is expandable as
        \begin{equation}
            \hat a^ \dagger _{\wpf} =\int dv   \,  \wpf (v) \hat a^\dagger_{v}\, ,
        \end{equation}
        where for spectral and temporal photon wavepacket, variable $v$ is frequency $\omega$ and time~$t$, respectively, and for a one-dimensional wavefront, is distance~$x$.  
        Figure~\ref{fig:pwp} illustrates a one-dimensional photon wavepacket~$\wpf(x)$ and a two-dimensional photon-wavepacket~$\wpf(x,y)$.
        At each coordinate, the height of the curves denote the photon-wavepacket's absolute value~$|\wpf|$, and the photon-wavepacket's phase~$\angle \wpf$ is color-coded.
        \par
        Let us consider the transformation of the pure state~\eqref{eqn:si=f(adag)} due to the unitary operator~$\hat{\text{U}}$, it gives
        \begin{equation}
          \begin{split}
            \hat{\text{U}}\lvert \psi \rangle &= \hat{\text{U}} f(\hat a_{\wpf}^ \dagger) \lvert 0 \rangle \\
            &=\sum_{n=0}^{\infty} c_n \frac{ \hat{\text{U}} \hat a_{\wpf}^{\dagger n} \hat{\text{U}}^{\dagger}}{\sqrt{n!}}\lvert 0 \rangle \\
            &=\sum_{n=0}^{\infty} c_n \frac{ \left(\hat{\text{U}} \hat a_{\wpf}^{\dagger} \hat{\text{U}}^{\dagger}\right)^n}{\sqrt{n!}}\lvert 0 \rangle\\
            &=f(\hat{\text{U}} \hat a_{\wpf}^{\dagger} \hat{\text{U}}^{\dagger}) \lvert 0 \rangle \\
            &=f(\hat a_{\wpf'}^{\dagger} ) \lvert 0 \rangle \, ,
          \end{split}
          \label{eqn:Usi}
        \end{equation}
        where equality~$\hat{\text{U}}^{\dagger} \lvert 0 \rangle=\lvert 0 \rangle$ and the unitarity property of operator~$\hat{\text{U}}^{\dagger}$($ \hat{\text{U}}^{\dagger} \hat{\text{U}} =\hat{\text{I}}$) is used, and the output single-photon  creation operator $\hat a_{\wpf'}^{\dagger}$ is defined as
        \begin{equation}
          \begin{split}
             \hat a_{\wpf'}^{\dagger}  &=\hat{\text{U}} \hat a_{\wpf}^{\dagger} \hat{\text{U}}^{\dagger}\, .
          \end{split}
          \label{eqn:si'}
        \end{equation}
        Equations~\eqref{eqn:Usi} and~\eqref{eqn:si'} denote that the unitary operator~$\hat{\text{U}}$ only changes the photon-wavepacket from $\wpf$ to $\wpf'$ but not the representation of the quantum state in the Fock space expressed by function~$f$ in Eq.~\eqref{eqn:si=f(adag)} and~\eqref{eqn:Usi}.
        Therefore, the only challenge to transform a quantum state due to a Fourier optical element is finding the transformation of photon-wavepacket creation operator equivalent to Eq.~\eqref{eqn:si'}.
        This simplification stems from the assumption that the Fourier optical elements do not possess nonlinear effects or photon-loss and therefore perform a unitary transformation on the quantum light.
        To make it short, our goal in this paper is to find a transformation similar to Eq.~\eqref{eqn:si'} for each Fourier optical element, which we use a full quantum mechanical approach to accomplish.
        \par
        Finally, let us highlight that extending the introduced evolution of quantum light through the Fourier optical systems to quantum mixed states and multi photon-wavepackets pure states~\cite{rezai_ieeeit_2021} is straightforward in the first approach.
        Similar to above, for Multiphoton-wavepacket quantum pure and mixed states,  only constituent photons' wavepackets get transformed while the photon statistics or Fock representation remains untouched.
        However, further simplification of such quantum states requires considering particular cases, which is beyond the scope of this paper.
        \subsection{Spatially Two Dimensional Quantum State of Light}        
        Equation~\eqref{eqn:si=f(adag)} assumes the photons of the pure light quantum state occupy identical modes~$\wpf$; identical spatial, spectral, and polarization modes.
        Quantum Fourier optics manipulate the spatial mode, and the manipulation usually depends on the spectrum and is independent of polarization.
        Therefore, as detailed in appendix~\ref{sec:csmco}, we drop the polarization and assume the quantum light is spectrally single-mode with frequency~$\moo$ and angular wavenumber~$\mko=\frac{\moo}{c}$, where $c$ is the speed of light.
        Furthermore, we assume the quantum light is propagating in $+z$-direction only (see Fig.~\ref{fig:sfa}).
        Therefore, the quantum light's photon-wavefunction can be associated with the wavefront function denoted as~$\wpf (x,y)$. 
        Since Fourier optical elements such as lenses would change the photon-wavefronts in $(x,y)$ domain, we add the photon-wavefront function~$\wpf=\wpf(x,y)$ as a subscript to the photon creation operators~$\hat a^ \dagger$ and introduce the photon-wavefront (photon-wavepacket) creation operator as~$\hat a^ \dagger _{\wpf}$, such that
        \begin{subequations}
          \begin{align}
            \label{eqn:axixy}
            \hat a^ \dagger _{\wpf} &=\iint dx dy   \,  \wpf (x,y) \hat a^\dagger_{x,y}\\
            \label{eqn:axikxky}
            &=\iint dk_x dk_y   \,  \wpfi (k_x,k_y) \hat a^\dagger_{k_x,k_y}\, ,
          \end{align}
          \label{eqn:axixykxky}
        \end{subequations}
      where creation operator $\hat a^\dagger_{x,y}$ creates a single photon at the point with coordinates $(x,y)$ associated with the
      two-dimensional continuous mode single-photon quantum state $\hat a^\dagger_{x,y} \lvert 0 \rangle= \lvert 1 \rangle_{x,y}\equiv\lvert x,y \rangle$.
      Correspondingly, in Eq.~\eqref{eqn:axixy}, the probability amplitude function of photon-wavepacket creation operator~$\hat a^ \dagger _{\wpf}$ is denoted by the photon-wavefront~$\wpf (x,y)$.
      The photon-wavepacket~$\wpfi (k_x,k_y)$ is the two-dimensional Fourier transform function of wavefront~ $\wpf (x,y)$, see Eq.~ \eqref{eqn:ftxi}, and corresponds to the single-photon probability amplitude at wavevector~$\mathbf{k}=\left(\mkx, \mky, \mkz=\sqrt{\mko^2-\mkx^2-\mky^2}\right)$, see Fig.~\ref{fig:sfa}.
      We use bold letters to represent wavevectors and non-bold letters for their wavenumber, $\left|\mathbf{\mko}\right|=\mko$. 
        \section{Evolution of a Quantum State and Phase Shifting Operators}
        In this section, we study the evolution of the quantum state of light due to various parameters such as time, displacement, and Fourier optical elements. 
        For all practical purposes, we show that all of these evolutions are formalizable by phase-shifting operators.        
        The primary phase-shifting operator can be expressed as follows~\cite{leonhardt_book_1997}:
        \begin{equation}
          \hat{\text{U}}= e^{-i \theta \hat n} = e^{-i \theta  \hat a^{\dagger} \hat a} \, ,
          \label{eqn:U}
        \end{equation}
        where $\hat n =  \hat a^{\dagger} \hat a $ is the number operator.
        The phase-shifting operator~$ \hat{\text{U}}$, as its name implies, phase-shifts the amplitude of operator $\hat a^{\dagger}$ by the phase value of $-\theta$ when acts on operator~$\hat a^{\dagger}$, i.e.,       
        \begin{equation}
          \begin{split}
            \hat{\text{U}} \hat a^\dagger \hat{\text{U}}^ \dagger  &=  \hat a ^{\dagger}  e^{-i \theta }\, .
          \end{split}
          \label{eqn:UadUd}
        \end{equation}
        \subsection{Time Evolution Operator}
        \label{sec:teo}
        The optical time evolution operator is
        \begin{equation}
          \begin{split}
            \hat{\text{V}}(t)&= e^{-i \hat H t/\hbar}\\
            &= e^{-i t\, \sum_{\mathbf{k}} \omega(k) \hat a^{\dagger}_{\mathbf{k}} \hat a_{\mathbf{k}}}\\
            &= e^{-i  \sum_{\mathbf{k}} \theta_k \hat n_{\mathbf{k}}} \\
            &= \prod_{\mathbf{k}}e^{-i  \theta_k \hat n_{\mathbf{k}}}\,
            \end{split}
          \label{eqn:Utgen}
        \end{equation}
        where $\hat H=\sum_{\mathbf{k}}\hbar \omega(k) \hat a^{\dagger}_{\mathbf{k}} \hat a_{\mathbf{k}}$ is the electromagnetic radiation Hamiltonian, and $\theta_{k} =\omega(k) t$, and  $\hat n_{\mathbf{k}} =  \hat a_{\mathbf{k}}^{\dagger} \hat a_{\mathbf{k}} $ is the number operator for the state with wavevector~${\mathbf{k}}$.
        Time evolution operator~\eqref{eqn:Utgen}, similar to Eq.~\eqref{eqn:U}, is a phase-shifting operator that, at time~$t$, provides wavevectors creation operator~$\hat a_{\mathbf{\mko}}^{\dagger}$ with phase shift~$-\theta_k=-\omega(k) t$, i.e., $ \hat{\text{V}}(t) \hat a^\dagger_{\mathbf{\mko}} \hat{\text{V}}^\dagger(t) =\hat a^\dagger_{\mathbf{\mko}}  e^{-i \omega(k) t}  $.
        \par
        This paper mainly considers spectrally single-mode quantum light.
        Single angular frequency~$\moo(k)= c|\mathbf{k}|= c k$ indicates that summation over the wavevector~$\mathbf{k}$ of Eq.~\eqref{eqn:Utgen} reduces to the summation over the wavevectors ending to the surface of the sphere with radius~$k$ (see ~Fig.~\ref{fig:sfa}).
        Consequently, single frequency photon-wavepacket creation operator~\eqref{eqn:axixykxky} gets a constant phase shift, $\hat a^ \dagger _{\wpf} \rightarrow  \hat a^ \dagger _{\wpf} e^{-i \moo t}$.
        This transformation is equivalent to wavefront transition from ${\wpf (x,y)}$ to ${\wpf(x,y)} \,   e^{-i \moo t}$.
        Thus the photon-wavefront only gets a constant phase factor~$e^{-i \moo t}$ over time without any overall changes in its shape.
        Therefore, without loss of generality, in this paper, we ignore the time evolution.
        One can show that the time evolution operator, as its name denotes, time-shifts the photon-wavepacket of spectrally multimode quantum light
         \begin{equation}
            \begin{split}
              \hat{\text{V}}(\tau) \lvert \psi \rangle &= \hat{\text{V}}(\tau)  f(\hat a^ \dagger_{\wpf}) \lvert 0 \rangle \\
              &=   f\left(\hat{\text{V}}(\tau) \hat a_{\wpf}^ \dagger\hat{\text{V}}^{\dagger}(\tau) \right) \lvert 0 \rangle  \\
              &=   f(\hat a^ \dagger_{\wpf'}) \lvert 0 \rangle 
            \end{split}
          \end{equation}
          where $\wpf'(\omega,x,y)=\wpf(\omega,x,y)e^{-i \omega \tau}$ and its Fourier transform shows that the photon-wavepacket in the time domain is shifted by time~$\tau$, i.e., $\wpft'(x,y,t)=\wpft(x,y,t+\tau)$ (See Eq.~\eqref{eqn:3dftxi} of appendix~\ref{sec:mfm}, which gives more detail about spectrally multimode photon-wavepacket, which is not the main focus of this paper.      
          \subsection{Displacement Transition Operator}
          \label{sec:dto}
        The displacement transition (evolution) operator in the $+z$-direction can be expressed as follows:      
        \begin{equation}
          \begin{split}
            \hat{\text{T}}(z) &= e^{+i z \sum_{\mathbf{k}} k_z \hat a_{\mathbf{k}}^{\dagger} \hat a_{\mathbf{k}}} \\
            &= \prod_{\mathbf{k}} e^{+i z k_z \hat a_{\mathbf{k}}^{\dagger} \hat a_{\mathbf{k}}}\\
            &= \prod_{\mathbf{k}} e^{+i z k_z \hat n_{\mathbf{k}}}\, ,
            \end{split}
          \label{eqn:Tz}
        \end{equation}
        and $k_z$ is the corresponding $z$~component of wavevector~$\mathbf{k}$.
        The displacement transition operator~\eqref{eqn:Tz} is also a phase-shifting type operator, which provides operator~$\hat a_{\mathbf{\mko}}^{\dagger}$ with phase shift~$\theta=-\mkz z$ for a displacement by~$z$:
        \begin{equation}
          \begin{split}
            \hat{\text{T}}(z) \hat a^\dagger_{\mathbf{\mko'}} \hat{\text{T}}^\dagger(z) &= \left( \prod_{\mathbf{k}} e^{+i z k_z \hat n_{\mathbf{k}}} \right)  \hat a^\dagger_{\mathbf{\mko'}}
             \left(\prod_{\mathbf{k}} e^{-i z k_z \hat n_{\mathbf{k}}}\right)\\
            &= e^{+i z  \mkz' \hat n_{\mathbf{\mko'}}}   \hat a^\dagger_{\mathbf{\mko'}} e^{-i z  \mkz' \hat n_{\mathbf{\mko'}}}\\
            &=\hat{\text{T}}_{\mathbf{\mko'}}(z) \hat a^\dagger_{\mathbf{\mko'}} \hat{\text{T}}_{\mathbf{\mko'}}^\dagger(z)\\
            &=  \hat a^\dagger_{\mathbf{\mko'}} e^{i \mkz' z} \, ,
          \end{split}
          \label{eqn:TzadTz}
        \end{equation}
      where $\hat{\text{T}}_{\mathbf{\mko}}(z)= e^{+i z  \mkz \hat n_{\mathbf{\mko}}}= e^{+i z  \mkz \hat a_{\mathbf{\mko}}^{\dagger} \hat a_{\mathbf{\mko}}} $ is the displacement transition operator for wavevector~$\mathbf{\mko}$.
      \subsubsection{\textbf{Fresnel Diffraction}}
        Under the paraxial~(Fresnel) approximation, propagating quantum electromagnetic wave makes a slight angle relative to the $z$-direction; therefore, $\mkz$ is reduceable to        
        \begin{equation}
          \mkz=\sqrt{\mko^2-\mkx^2-\mky^2}\approx \mko \left(1-\frac{\mkx^2}{2\mko^2}-\frac{\mky^2}{2\mko^2}\right)\, .
          \label{eqn:papp}
        \end{equation}
      Therefor, the Fresnel displacement transition quantum operator is
      
        \begin{equation}
          \begin{split}
            \hat{\text{T}}^{_{\text{Fr}}}(z) 
            &= \prod_{\mathbf{k}} e^{+i z\, k_z  \hat n_{\mathbf{k}}}\\
            &= \prod_{\mathbf{k}} e^{+i z \, \mko \left(1-\frac{\mkx^2}{2\mko^2}-\frac{\mky^2}{2\mko^2}\right) \hat n_{\mathbf{k}}}\\
            &= \prod_{\mathbf{k}} \hat{\text{T}}^{_{\text{Fr}}}_{\mathbf{\mko}}(z) \, ,
            \end{split}
          \label{eqn:Tzfr}
        \end{equation}
      where $\hat{\text{T}}^{_{\text{Fr}}}_{\mathbf{\mko}}(z)=e^{+i z \, \mko \left(1-\frac{\mkx^2}{2\mko^2}-\frac{\mky^2}{2\mko^2}\right)\hat n_{\mathbf{k}}}$ is the Fresnel displacement transition operator for wavevector~$\mathbf{\mko}$.
           
        Fresnel approximation is used in the following to simplify the displacement operation on the quantum state of light.
        Regarding the discussions about Eq.~\eqref{eqn:Usi} and~\eqref{eqn:si'}, to obtain the operation of the unitary Fresnel displacement transition~\eqref{eqn:Tzfr} on a quantum state of light, we need to transform the photon-wavepacket creation operator; using the same procedure as Eq.~\eqref{eqn:TzadTz}, we get
      \begin{equation}
           \begin{split}
             \hat{\text{T}}^{_{\text{Fr}}}(z) \hat a^ \dagger _{\wpf}\hat{\text{T}}^{_{\text{Fr}}\dagger}&(z)
            =\iint dk_x dk_y   \,  \wpfi (k_x,k_y)  \hat{\text{T}}^{_{\text{Fr}}}(z) \hat a^\dagger_{k_x,k_y} \hat{\text{T}}^{_{\text{Fr}}\dagger}(z)\\
            &=\iint dk_x dk_y   \,  \wpfi (k_x,k_y)  e^{i \mko \left(1-\frac{\mkx^2}{2\mko^2}-\frac{\mky^2}{2\mko^2}\right)z } \hat a^\dagger_{k_x,k_y} 
            \\
            &=\iint dk_x dk_y   \,  \wpfi (k_x,k_y)   \tilde {h}^{_{\text{Fr}}}_{z}(k_x,k_y) \hat a^\dagger_{k_x,k_y} 
            \\
            &=\iint dk_x dk_y   \,   \wpfi^{_{\text{Fr}}}_{z}(k_x,k_y)  \hat a^\dagger_{k_x,k_y} 
            \\
            &= \iint dx dy   \, \wpf^{_{\text{Fr}}}_{z}(x,y) \,   \hat a^\dagger_{x,y}\\
            &= \hat a^ \dagger _{\wpf^{_{\text{Fr}}}_{z}}\, ,
          \end{split}
          \label{eqn:tatz}
        \end{equation}        
        where Fresnel transfer function~$\tilde {h}^{_{\text{Fr}}}_{z}(k_x,k_y)$ is defined as
        \begin{equation}
          \begin{split}
            \tilde{h}^{_{\text{Fr}}}_{z}(k_x,k_y)&=  e^{i \mko \left(1-\frac{\mkx^2}{2\mko^2}-\frac{\mky^2}{2\mko^2}\right)z }\, ,   
          \end{split}
          \label{eqn:thfr}
        \end{equation}
        and the Fresnel transformed photon-wavepacket~$\wpfi^{_{\text{Fr}}}_{z}(k_x,k_y)$  of the input photon-wavepackt~$\wpfi (k_x,k_y)$ for a displacement by~$z$ is
        \begin{equation}
          \begin{split}
            \wpfi^{_{\text{Fr}}}_{z}(k_x,k_y) &= \wpfi (k_x,k_y)  \tilde{h}^{_{\text{Fr}}}_{z}(k_x,k_y) \\
            &= \wpfi (k_x,k_y)  e^{i \mko \left(1-\frac{\mkx^2}{2\mko^2}-\frac{\mky^2}{2\mko^2}\right)z }\, ,
          \end{split}
          \label{eqn:frxikk}
          \end{equation}
          which is defined in the wavevector domain.
          For the spatial domain, use Eq~\eqref{eqn:iftxi} and 
          the convolution theorem, which recast Fresnel transformed wavepacket~\eqref{eqn:frxikk} into the following form
          \begin{equation}
            \begin{split}
              \wpf^{_{\text{Fr}}}_{z} (x,y)& =\frac{1}{2 \pi} \iint dk_x dk_y \,  \wpfi^{_{\text{Fr}}}_{z} (k_x,k_y) e^{i\left(k_x x +k_y y \right)} \\
              & =\frac{1}{2 \pi} \iint dk_x dk_y \,  \wpfi (k_x,k_y)  \tilde{h}^{_{\text{Fr}}}_{z}(k_x,k_y)e^{i\left(k_x x +k_y y \right)}\\            
              & = \wpf (x,y)\ast {h}^{_{\text{Fr}}}_{z} (x,y)\, ,
            \end{split}
            \label{eqn:frxixy}
          \end{equation}
          where symbol~$\ast$ denotes the convolution operation defined in Eq.~\eqref{eqn:convdef}.
          Equation~\eqref{eqn:frxixy} is known as the Fresnel diffraction formula.      
          The Fresnel impulse response of free space propagation~$ {h}^{_{\text{Fr}}}_{z} (x,y)$ is given by equation
        \begin{equation}
          \begin{split}
            {h}^{_{\text{Fr}}}_{z} (x,y)& =\frac{1}{2 \pi} \iint dk_x dk_y \, \tilde{h}^{_{\text{Fr}}}_{z}(k_x,k_y) e^{i\left(k_x x +k_y y \right)} 
            \\
            & =\frac{-i \mko}{z} e^{i \mko z} e^{\frac{i \mko}{2z} ( x^2+y^2)}  \, .
          \end{split}
          \label{eqn:hfr}
        \end{equation}
    \par
    Considering equations~\eqref{eqn:tatz} and \eqref{eqn:frxixy}, one may note that Fresnel diffraction performs a single-photon quantum state convolution with the Fresnel impulse response function~\eqref{eqn:hfr}.
    For more detail about the quantum convolution, see section~\ref{sec:quconv} on the 4f-processor.

    \subsubsection{\textbf{Fraunhofer Diffraction}}
    The Fraunhofer diffraction or the far-field diffraction of photon-wavepacket creation operator~$\hat a^ \dagger _{\wpf}$  is the limiting case of the Fresnel diffraction~\eqref{eqn:tatz} where the displacement~$z$ is much bigger than the size (width) of photon-wavefront~${\wpf (x,y)}$, denoted by~$\Delta x, \Delta y$.
    Mathematically speaking, if $\wpf (x,y)\approx 0 \ \forall\ x \notin (-\frac{\Delta x}{2}, \frac{\Delta x}{2} ) \ \cup \ y \notin (-\frac{\Delta y}{2},\frac{\Delta y}{2}) $, at distance
    \begin{equation}
      z \gg \frac{\mko}{8} ( \Delta x^2+ \Delta y^2)
    \end{equation}
    the Fresnel wavefront~\eqref{eqn:frxixy} reduces to
    \begin{equation}
          \begin{split}
            \wpf^{_{\text{Fr}}}_{z} (x,y)& = \wpf (x,y)\ast {h}^{_{\text{Fr}}}_{z} (x,y)\\
            & =\frac{1}{2 \pi} \iint  dx' dy' \,  \wpf (x',y')\, {h}^{_{\text{Fr}}}_{z} (x-x',y-y')\\
            & \approx  \frac{-i \mko}{z}   e^{i \mko z} e^{\frac{i \mko}{2z} (x^2+y^2)} \\
            &\qquad \times  \frac{1}{2 \pi} \iint  dx' dy' \,  \wpf (x',y')\, e^{\frac{-i \mko}{z} ( xx'+yy')}\\
            & =  \frac{-i \mko}{z}   e^{i \mko z} e^{\frac{i \mko}{2z} (x^2+y^2)} \wpfi (\frac{\mko}{z}x,  \frac{\mko}{z}y)\\
            &= \wpf^{_{\text{Fh}}}_{z} (x,y)\, ,
          \end{split}
          \label{eqn:}
        \end{equation}
        where Eq.~\eqref{eqn:frxixy}, \eqref{eqn:hfr} and \eqref{eqn:ftxi} are used, and the Fraunhofer wavefront is defined as $ \wpf^{_{\text{Fh}}}_{z} (x,y)=\frac{-i \mko}{z}   e^{i \mko z} e^{\frac{i \mko}{2z} (x^2+y^2)} \wpfi (\frac{\mko}{z}x, \frac{\mko}{z}y)$, which is proportional to the Fourier transform of the input wavefront at point~$ (\frac{\mko}{z}x,\frac{\mko}{z}y)$.
        Consequently, the Fraunhofer diffraction of creation operator~$\hat a^ \dagger _{\wpf}$ is expressible as
        \begin{equation}
            \hat{\text{T}}^{_{\text{Fh}}}(z) \hat a^ \dagger _{\wpf}\hat{\text{T}}^{_{\text{Fh}}\dagger}(z) = \hat a^ \dagger _{\wpf^{_{\text{Fh}}}_{z}}
          \end{equation}
        denoting continuous mode quantum Fourier transformation on a single-photon number state.
        For more detail about the quantum Fourier transform, see section~\ref{sec:lensop} on the lens operator.
        \subsection{ Two-Dimensional Quantum Spatial Phase Modulation Operator}
          \label{sec:spmo}
        We introduce two-dimensional spatial phase modulation operator~$\hat{\text{U}}_{\phi}$ on coordinate~$x$ and~$y$ as follows:        
        \begin{equation}
          \hat{\text{U}}_{\phi}  = e^{- i \sum_{x,y} \phi(x,y) \hat n_{x,y}} \, ,
          \label{eqn:Uphixy}
        \end{equation}
        where $\hat n_{x,y} =  \hat a_{x,y}^{\dagger} \hat a_{x,y} $ is the number operator for position~$(x,y)$.
        Phase-shifting operator~\eqref{eqn:Uphixy} provides operator~$\hat a_{x,y}^{\dagger}$ with phase shift~$\theta=\phi(x,y)$:        
        \begin{equation}
          \begin{split}
            \hat{\text{U}}_{\phi} \hat a^\dagger_{x',y'} \hat{\text{U}}^\dagger_{\phi} &= e^{-i  \sum_{x,y} \phi(x,y) \hat n_{x,y}}   \hat a^\dagger_{x',y'} e^{i  \sum_{x,y} \phi(x,y)  \hat n_{x,y}}\\
            &= e^{-i   \phi(x',y')  \hat n_{x',y'}}   \hat a^\dagger_{x',y'} e^{i  \phi(x',y')  \hat n_{x',y'}}\\         
            &=  \hat a^\dagger_{x',y'} e^{-i \phi(x',y')} \, .
          \end{split}
          \label{eqn:uadxyu}
        \end{equation}
      To obtain Eq.~\eqref{eqn:uadxyu} from Eq.~\eqref{eqn:Uphixy}, we use the same argument as in Eq.~\eqref{eqn:Tz} to reach Eq.~\eqref{eqn:TzadTz}.
      Therefore, operator~$ \hat{\text{U}}_{\phi}$ transforms photon-wavepacket creation operator~$\hat a^ \dagger _{\wpf}$ as follows:
        \begin{equation}
          \begin{split}
            \hat{\text{U}}_{\phi} \hat a^ \dagger _{\wpf}\hat{\text{U}}^\dagger_{\phi} &=\iint dx dy   \,  \wpf (x,y)  \hat{\text{U}}_{\phi} \hat a^\dagger_{x,y} \hat{\text{U}}^\dagger_{\phi}\\
            &=\iint dx dy   \,  \wpf (x,y) e^{-i \phi(x,y)}  \hat a^\dagger_{x,y} \\
            &=\iint dx dy   \,  \wpf' (x,y)  \hat a^\dagger_{x,y} \\                        
            &=\hat a^ \dagger _{\wpf'}\, ,            
            \end{split}
          \label{eqn:UadU}
        \end{equation}
      where  $\wpf'(x,y) = \wpf(x,y)  e^{-i \phi(x,y)}$ and Eq.~\eqref{eqn:axixy} and \eqref{eqn:uadxyu} are used.
      \subsubsection{\textbf{Ideal Thin Lens Quantum Modulation}}
      
        An ideal thin lens is a two-dimensional spatial phase modulator meaning its spatial transformation phase-shifts the two-dimensional photon-wavepackets.
        The phase function~$\phi_l(x,y)$ of a classical and ideal lens with focal length $f$ is
        \begin{equation}
          \phi_l(x,y)=\frac{\mko}{2f} (x^2+y^2) \, .
          \label{eqn:phil}
        \end{equation}
        From Eq.~\eqref{eqn:Uphixy} and~\eqref{eqn:phil}, the lens phase-shifting operator~$\hat{\text{U}}_{\phi_l}$ can be expressed as follows       
        \begin{equation}
          \hat{\text{U}}_{\phi_l} = e^{- i \sum_{x,y} \frac{\mko}{2f} (x^2+y^2)\hat n_{x,y}} \, .
          \label{eqn:Ul}
        \end{equation}
      \subsubsection{\textbf{Diffraction Grating Quantum Modulation}}
      \label{sec:grating}
      \begin{figure}[!t]
          \centering
          \includegraphics[width=\columnwidth]{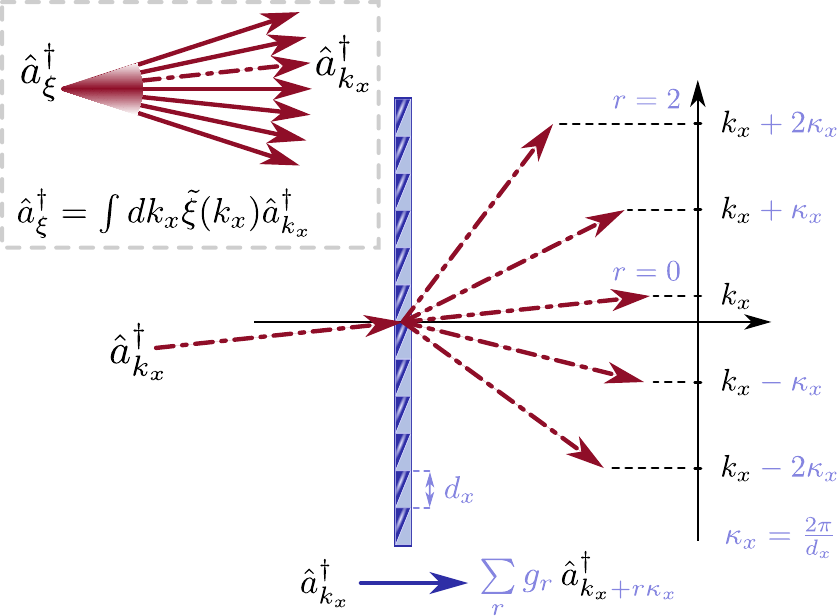}
          \caption{\textbf{Diffraction grating quantum modulation.}
            The inset demonstrates a photon-wavepacket in a superposition of wavevectors with amplitude $|\mathbf{k}|=k$, labeled with their~$x$-components.
            The grating shifts each wavevector to a superposition of wavevectors, whose $x$-components are shifted by $r\kappa_x$, where $\kappa_x$ is the frequency spacing of the grating. 
          }\label{fig:grating}
        \end{figure}
       
      A diffraction grating is a periodic spatial phase modulator.
      Assume the periodic structure of the grating is in the $x$-direction.
      In other words, assume the grating phase function is~$\phi_g(x)$ with period $\xg$.
      Therefore, the corresponding grating phase factor~$e^{-i \phi_g(x)}$ is periodic with spatial frequency $\kappa_x=\frac{2 \pi }{\xg}$, according to the Fourier transform theory, expandable as 
      \begin{equation}
        \begin{split}
          e^{-i \phi_g(x)} &=\sum_r \gr_{r} e^{i\left( r\kappa_x x\right)}\\
          &=\sum_r \gr_{r} e^{i\left( \frac{ 2\pi r}{\xg} x\right)}\, ,
        \end{split}
        \label{eqn:ft[ephig]}
      \end{equation}
      where $ \gr_{r}$ indicates the $r$th Fourier coefficient of the grating phase factor.
      Therefore, as Eq.~\eqref{eqn:UadU} denotes, the diffraction grating transforms photon-wavepacket creation operator~$\hat a^{\dagger }_{\wpf}$ as follows        
        \begin{equation}
          \begin{split}
            \hat{\text{U}}_{\phi_g} \hat a^ \dagger _{\wpf}\hat{\text{U}}^\dagger_{\phi_g}             &=\iint dx dy   \,  \wpf (x,y) e^{-i \phi_g(x)}  \hat a^\dagger_{x,y}\\    
            &=\sum_r\gr_{r} \iint dx dy   \,  \wpf (x,y)  e^{i r\kappa_x x }
            \hat a^\dagger_{x,y} \\
            &=\sum_r\gr_{r} \iint dx dy   \,  \wpf_{r} (x,y) 
            \hat a^\dagger_{x,y}\\
            &=\sum_r\gr_{r} \iint dk_x dk_y   \,  \wpfi_{r} (k_x,k_y) 
            \hat a^\dagger_{k_x,k_y} 
            \end{split}
          \label{eqn:UgadUg0}
        \end{equation}
        where the $r$th diffraction order photon-wavefront is defined as $\wpf_{r}(x,y)= \wpf (x,y)  e^{i r\kappa_x x }$ and its inverse Fourier transform (Eq.~\eqref{eqn:ftxi}) gives its corresponding representation in the wavevector space, that is
        \begin{equation}
          \begin{split}
            \wpfi_r (k_x,k_y)& =\frac{1}{2 \pi} \iint dx dy \,  \wpf_r (x,y) e^{-i\left(k_x x +k_y y\right) } \\
            & =\frac{1}{2 \pi} \iint dx dy \,  \wpf (x,y) e^{-i\left((k_x-r\kappa_x) x +k_y y\right) } \\
            &= \wpfi (k_x-r\kappa_x,k_y)\, .
          \end{split}
          \label{eqn:iftxir}
        \end{equation}
        Substituting Eq.~\eqref{eqn:iftxir} in Eq.~\eqref{eqn:UgadUg0} gives
        \begin{equation}
          \begin{split}
            \hat{\text{U}}_{\phi_g} \hat a^ \dagger _{\wpf}\hat{\text{U}}^\dagger_{\phi_g}            
            &=\sum_r\gr_{r} \iint dk_x dk_y   \,  \wpfi (k_x-r\kappa_x,k_y) 
            \hat a^\dagger_{k_x,k_y} \\
            &=\sum_r\gr_{r} \iint dk_x dk_y   \,  \wpfi (k_x,k_y) 
            \hat a^\dagger_{k_x+r\kappa_x,k_y}\, .
            \end{split}
          \label{eqn:UgadUg1}
        \end{equation}
        Comparing Eq.~\eqref{eqn:UgadUg1} with Eq.\eqref{eqn:axikxky} indicates  the diffraction grating performs the following quantum transformation in the wavevector space        
        \begin{equation}
          \begin{split}
              \hat{\text{U}}_{\phi_g}  \hat a^\dagger_{k_x,k_y}\hat{\text{U}}^\dagger_{\phi_g}= \sum_r \gr_{r} \hat a^\dagger_{k_x+r\kappa_x,k_y}\, .
            \end{split}
          \label{eqn:UgadkUg}
        \end{equation}
        The quantum grating transformation equation~\eqref{eqn:UgadkUg} indicates that at the~$r$th diffraction order, the diffraction amplitude equals the $r$th Fourier coefficient of the grating $\gr_{r}$.
        Furthermore, the incident wavevector~$\mathbf{k}$ with $x$ component~$k_x$ transforms to the wavevector~$\mathbf{k}^{r}$ with $x$ component~$k^r_x=k_x+r\kappa_x$.
        Assume the angle of the incident wavevector~$\mathbf{k}$ and the~$r$th diffracted wavevector~$\mathbf{k}^{r}$ with the grating are~$\theta_i$ and ~$\theta_r$, respectively.
  	Thus, the transformation equation $k^r_x=k_x+r\kappa_x$, indicated by Eq.~\eqref{eqn:UgadkUg}, reduces to the well-known grating equation:
        \begin{equation}
          \begin{split}
             \sin(\theta_r)&=\sin(\theta_i)+ r\frac{\lambda}{\xg}\, ,
            \end{split}
          \label{eqn:geq}
        \end{equation}
        where $\lambda=\frac{2\pi}{\mko}$ denotes the photon wavelength.
        \par
        One can see in this one-dimensional modulation of the grating on the $x$-components of the quantum light; we could ignore the $y$-component from the consideration, as argued in appendix~\ref{sec:csmco} and shown in Fig.~\ref{fig:grating}.  
      \section{Lens Operator as Quantum Optical Fourier Transformer}
      \label{sec:lensop}
        \begin{figure}[!t]
          \centering
          \includegraphics[width=\columnwidth]{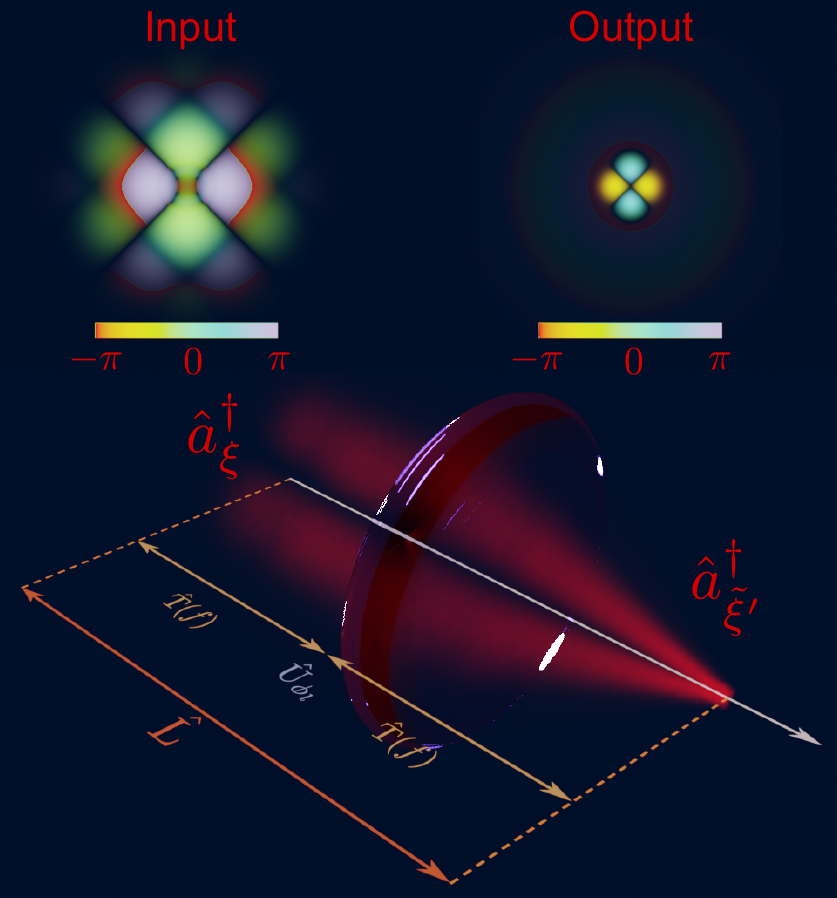}
          \caption{\textbf{Lens quantum operation.}
            The input of the lens is considered to be the quantum light in its front focal plane, and the output is the quantum light in its back focal plane.
            Therefore, the lens operator $\hat{\text{L}}$ is a sequence of three transformations: The displacement transition up to the lens~$\hat{\text{T}}(f)$, lens spatial phase modulation~$\hat{\text{U}}_{\phi_l}$, and final displacement transition to the back focal plane~$\hat{\text{T}}(f)$.
            The figure, moreover, shows simulation results: the input ($\wpf(x,y)$) and the output ($\wpfi'(x,y)$) photon-wavepackets at the top and the propagation throughout the lens operator ($\lvert \wpf^{_{\text{Fr}}}_{z}(x,y)\rvert^2$). The color of the input and output photon-wavepackets represent the phase of photon-wavepackets, and the color’s intensity reflects the amplitude of photon-wavepackets.
            At each point of propagation through the lens system, the intensity of the red light shows the corresponding single-photon probability ($\lvert \wpf^{_{\text{Fr}}}_{z}(x,y)\rvert^2$).
          }\label{fig:lens}
        \end{figure}
        Consider a quantum light impinging upon a lens; consider its quantum state at the front focal plane and the back focal plane, respectively, as the lens's input and output.
        We express the quantum transformation of such a passive optical unit with operator~$\hat{\text{L}}$.
        \par
        Fig.~\ref{fig:lens} shows the transformation of the lens operator~$\hat{\text{L}}$ on a quantum light.
        The lens transformation is structured as a sequence of three quantum transformations: the quantum displacement transformation to the frontend of the lens~$\hat{\text{T}}(f)$, the quantum modulation from the front to the backend of the lens~$\hat{\text{U}}_{\phi_l}$, followed by the second quantum displacement to the back focal plane~$\hat{\text{T}}(f)$; mathematically speaking
        \begin{equation}
          \hat{\text{L}} =\hat{\text{T}}(f)\hat{\text{U}}_{\phi_l}  \hat{\text{T}}(f)\, .
          \end{equation}
        \par
        Appendix~\ref{sec:lens} shows that the effect of the lens operator~$\hat{\text{L}}$ on a photon-wavepacket creation operator~$\hat a^{\dagger }_{\wpf}$ is to reshape its wavefront, the quantum reshaping operation is as follows (Eq.~\eqref{eqn:LadLx}): 
        \begin{equation}
          \begin{split}
             \hat{\text{L}}  \hat a^ \dagger _{\wpf} \hat{\text{L}}^{\dagger} 
             &= \hat a^ \dagger _{\wpfi'}\, .
           \end{split}
           \label{eqn:LadLmt}
         \end{equation}
         From appendix~\ref{sec:lens}, Eq.~\eqref{eqn:wpfi'(x,y)}, which assumes the Fresnel approximation for the transition operator, $\hat{\text{T}}(z) =\hat{\text{T}}^{_{\text{Fr}}}(z)$, the lens reshapes the wavefront from~$\wpf$ to $\wpfi'$.
         The output photon wavefront~$\wpfi'$ corresponds to the input wavefront's Fourier transform~$\wpfi$ rescaled with factor~$\frac{\mko}{f}$ as
         \begin{equation}
          \begin{split}
            \wpfi' (x,y)  &= \frac{-i\mko}{f}  e^{2 i \mko f}  \wpfi ( \frac{\mko}{f}x,  \frac{\mko}{f}y )\, .
             \end{split}
          \label{eqn:wpfi'(x,y)main}
        \end{equation}
         \par
         As a general example, consider the generic class of pure quantum state of identical photons with photon-wavefront~$\wpf$ (Eq.~\eqref{eqn:si=f(adag)}, $\lvert \psi \rangle =f(\hat a^ \dagger _{\wpf}) \lvert 0 \rangle$) as the inputs of the lens operator~$\hat{\text{L}}$.
        The lens operator~$\hat{\text{L}}$ transforms such a quantum light as follows
      \begin{equation}
        \begin{split}
          \lvert \Psi \rangle &= \hat{\text{L}} \lvert \psi \rangle\\
          &=\hat{\text{L}} f(\hat a^ \dagger _{\wpf}) \lvert 0 \rangle\\
          &= \sum^{\infty}_{n=0} \frac{c_n}{\sqrt{n!}} \hat{\text{L}} \hat a^{\dagger n}_{\wpf}\lvert 0 \rangle
          \\
          &= \sum^{\infty}_{n=0} \frac{c_n}{\sqrt{n!}} \left( \hat{\text{L}} \hat a^{\dagger}_{\wpf} \hat{\text{L}}^{\dagger} \right)^n \hat{\text{L}}\lvert 0 \rangle
          \\
          &= \sum^{\infty}_{n=0} \frac{c_n}{\sqrt{n!}} \hat a^{\dagger n} _{\wpfi'} \lvert 0 \rangle\\
          &=f(\hat a^ \dagger _{\wpfi'}) \lvert 0 \rangle\, ,
        \end{split}
        \label{eqn:Lpsi}
        \end{equation}
        where the unitarity of the lens operator ($\hat{\text{L}}^{\dagger}\hat{\text{L}} =\hat{\text{I}}$),  equation~\eqref{eqn:LadLmt} and invariance of the vacumm by the lens operation ($\hat{\text{L}}\lvert 0 \rangle=\lvert 0 \rangle$) are used.
        \par        
        \begin{itemize}
        \item \textit{ Example 1: Single-photon input}
        \par
        Consider the particular case of single-photon number state~$\lvert 1 \rangle_{\wpf}= \hat a^{\dagger}_{\wpf}\lvert 0 \rangle$ as the input of the lens operator, the output of the lens would be the following quantum state
        \begin{equation}
          \begin{split}
            \hat{\text{L}} \lvert 1 \rangle_{\wpf}&= \hat{\text{L}} \hat a^{\dagger}_{\wpf}\lvert 0 \rangle =\hat{\text{L}} \hat a^{\dagger}_{\wpf}\hat{\text{L}}^{\dagger}\lvert 0 \rangle=\hat a^{\dagger}_{\wpfi'}\lvert 0 \rangle \\
            &=\lvert 1 \rangle_{\wpfi'}\, .\\
          \end{split}
        \end{equation}
        Equations~\eqref{eqn:LadLx} and \eqref{eqn:wpfi'(x,y)} give the representation of the above state in the spatial space as follows
        \begin{equation}
           \begin{split}
              \hat{\text{L}} \lvert 1 \rangle_{\wpf} 
              &=   \iint dx dy \,  \wpfi' (x, y )\, \hat a^\dagger_{x,y} \lvert 0 \rangle\\
              &=    \frac{-i\mko}{f}  e^{2 i \mko f}   \iint dx dy \, \wpfi ( \frac{\mko}{f}x,  \frac{\mko}{f}y )\, \hat a^\dagger_{x,y} \lvert 0 \rangle\\
              &=    -i\frac{ f}{\mko}  e^{2 i \mko f}   \iint du dv \, \wpfi (u, v )\,  \hat a^\dagger_{\frac{f}{\mko} u,\frac{f}{\mko} v}\lvert 0 \rangle\\
              &=    -i e^{2 i \mko f}   \iint du dv \, \wpfi (u, v )\,  \hat a^\dagger_{u,v}\lvert 0 \rangle\\
              &=    -i e^{2 i \mko f}   \iint du dv \, \wpfi (u, v )\, \lvert 1 \rangle_{u,v}\, ,
           \end{split}
           \label{eqn:L|1>}
         \end{equation}
         where space linear rescaling relations $ u=\frac{\mko}{f}x\, , v=\frac{\mko}{f}y$ and the field rescaling proportionality relation~\eqref{eqn:propforad}, $\hat a^{\dagger}_{\alpha u, \beta v}=\frac{1}{\sqrt{\lvert \alpha \beta\rvert }}\hat a^{\dagger}_{u,v}$, are used.
         As Eq.~\eqref{eqn:ftxi} shows, $\wpfi(u,v)$ is the Fourier transform of $\wpf(x,y)$.
         Therefore,  ignoring the trivial phase factor~$-i e^{2 i \mko f}$, equation~\eqref{eqn:L|1>} denotes that a lens makes continuous mode quantum Fourier transformation on a single-photon number state.
         In other words, the wave function (wavefront) of a single-photon at the output of a lens unit is the Fourier transform of the single-photon's wave function at the input of the lens.
         \item \textit{ Example 2: Glauber State input}
        \par
        If the input of the Lens is a Glauber state, i.e., 
        $\lvert\alpha \rangle_{\wpf}=\exp( -|\alpha|^ 2/2 + \alpha \hat a^ \dagger _{\wpf})\lvert 0 \rangle$,
        the lens operation gives another Glauber state with the same amplitude~$\alpha$:
        \begin{equation}
          \begin{split}
            \hat{\text{L}} \lvert \alpha \rangle_{\wpf}&=\hat{\text{L}}e^{-|\alpha|^ 2/2 + \alpha \hat a^ \dagger _{\wpf}} \hat{\text{L}}^{\dagger}\lvert 0 \rangle\\
            &=e^{ -|\alpha|^ 2/2 + \alpha \hat a^ \dagger _{\wpfi'}}\lvert 0 \rangle\\
            &=\lvert\alpha \rangle_{\wpfi'}\, .
          \end{split}
        \end{equation}
        where also $\hat{\text{L}}^{\dagger} \lvert 0 \rangle=\lvert 0 \rangle$.
      \end{itemize}      
      \section{Quantum Convolution via a 4\lowercase{f} Quantum Signal Processing System}
      \label{sec:quconv}
      \begin{figure}[!t]
          \centering
          \includegraphics[width=\columnwidth]{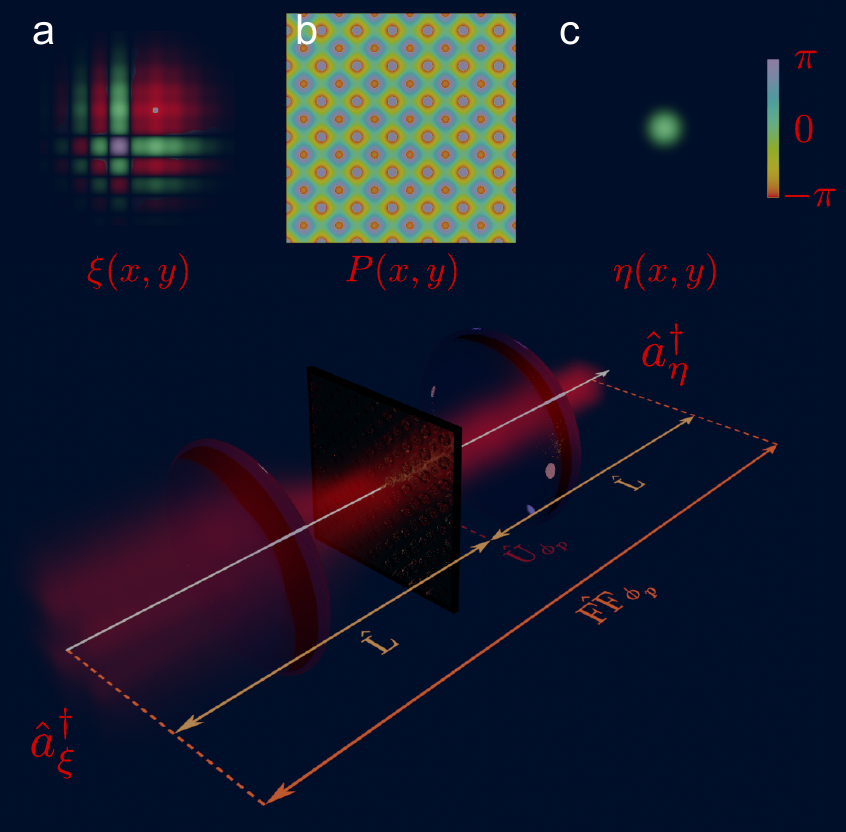}
          \caption{\textbf{4f Quantum Signal Processing System}
            comprises two similar lenses that in their confocal plane there is a pupil, a spatial phase modulator, associated with the intended quantum processing.
            \textbf{a} shows the input photon-wavepacket~$\wpf(x,y)$. The phase of photon-wavepacket~$\wpf(x,y)$ is color-coded according to the color bar at the upper-right corner of the figure.
            The strength of the color reflects the amplitude of photon-wavepacket~$\lvert \wpf(x,y) \rvert$.
            \textbf{b} shows the pupil function~$\Pf(x,y)$. Similar to \textbf{a}, the phase and the amplitude are displayed with color and the intensity of the color. Since the pupil is phase-only, the intensity of the colors remains constant and corresponds to unity, $\lvert \Pf(x,y)\rvert =1$.
            \textbf{c} is like  \textbf{a} but for output photon wavepacket $\wpfo(x,y)$.
            In each point of the 4f-system, the intensity of red light demonstrates the probability ($\lvert \wpf^{_{\text{Fr}}}_{z}(x,y)\rvert^2$) of the propagating photon-wavepacket.
          }\label{fig:4f}
        \end{figure}
        Figure~\ref{fig:4f} shows a 4f-processing system consisting of two identical lenses of confocal length~$f$, with a spatial phase modulator of pupil function~$\phi_p(x,y)$ located at the confocal plane (the Fourier plane of the primary lens).
        The corresponding quantum operator of this 4f-processing system is expressible as
        \begin{equation}
          \hat{\text{FF}}_{\phi_p}= \hat{\text{L}}  \hat{\text{U}}_{\phi_p} \hat{\text{L}}\, .
          \end{equation}
        As in Eq.~\eqref{eqn:LadLmt}, the first lens transforms wavepacket creation operator~$\hat a^ \dagger _{\wpf}$ to $\hat a^ \dagger _{\wpfi'}$ ($\hat{\text{L}}  \hat a^ \dagger _{\wpf} \hat{\text{L}}^{\dagger} = \hat a^ \dagger _{\wpfi'}$),
        where, as shown in Eq.~\eqref{eqn:wpfi'(x,y)}, $\wpfi'$ corresponds to the Fourier transform of the input wavefront~${\wpf}$.

        Afterward, the phase-only pupil~$\Pf(x,y)= e^{-i \phi_p(x,y)}$ transforms the operator~$ \hat a^ \dagger _{\wpfi'}$ to $ \hat a^ \dagger _{\wpfoi'}$ ($\hat{\text{U}}_{\phi_p} \hat a^ \dagger _{\wpfi'}\hat{\text{U}}^\dagger_{\phi_p} =\hat a^ \dagger _{\wpfoi'}$), where according to Eq.~\eqref{eqn:UadU}, the photon-wavefront at the backend of the pupil ($\wpfoi'$) is
        \begin{equation}
          \begin{split}
            \wpfoi'(x,y)&=\wpfi'(x,y) \Pf(x,y)\\
            &=\wpfi'(x,y) e^{-i \phi_p(x,y)} \, .
            \end{split}
          \label{eqn:wpfi''}
        \end{equation}
        Finally, the second lens transforms the operator~$\hat a^ \dagger _{\wpfoi'}$ to $\hat a^ \dagger _{\wpfo}$  ($\hat{\text{L}}  \hat a^ \dagger _{\wpfoi'} \hat{\text{L}}^{\dagger} = \hat a^ \dagger _{\wpfo}$), where, similar to the first lens, the output photon-wavefront~$\wpfo$ corresponds to the Fourier transform of the second lens's input photon-wavefront~$\wpfoi'$, and,  as Eq.~\eqref{eqn:4f} shows, it becomes
        \begin{equation}
           \begin{split}
             \wpfo(x,y)
             &=  \wpf (-x,-y)\ast h_{\hsub}(x,y)\, .
                \end{split}
          \label{eqn:wpfii'(x,y)}
        \end{equation}
        Equation~\eqref{eqn:wpfii'(x,y)} indicates that the output photon-wavefront~$\wpfo$ of the 4f-processing system is the convolution of its  impulse response~$h_{\hsub}(x,y)$ and the parity transformed input photon-wavefront~$\wpf (- x, - y)$.
        For the 4f-processing system with pupil~$\Pf(x,y)$, as equation~\eqref{eqn:hff}, $h_{\hsub}(x,y)= -  e^{i 4 \mko f} \frac{\mko^2}{f^2} \fPf(\frac{\mko}{f}x, \frac{\mko}{f} y)$, denotes, its impulse response~$ h_{\hsub}(x,y)$ corresponds to the rescaled Fourier transform of the pupil function~$\Pf(x,y)$ by factor~$\frac{k}{f}$.       
        In short, the 4f-operator~$\hat{\text{FF}}_{\phi_p}$ transforms its input photon-wavepacket creation operator~$ \hat a^ \dagger _{\wpf}$ as follows
        \begin{equation}
          \begin{split}
            \hat{\text{FF}}_{\phi_p}\, \hat a^ \dagger _{\wpf}\, \hat{\text{FF}}^{\dagger}_{\phi_p}&=\hat a^ \dagger _{\wpfo}\\
            &=\iint dx dy   \,  \wpfo (x,y) \,  \hat a^\dagger_{x,y}\\
            &=\iint dx dy   \,  \wpf (-x,-y)\ast h_{\hsub}(x,y) \, \hat a^\dagger_{x,y}\, .
            \end{split}
          \label{eqn:ffadff}
        \end{equation}
        \par
        Following the same procedure as Eq.~\eqref{eqn:Lpsi}, one can show that the unitary, linear, 4f-operator~$ \hat{\text{FF}}_{\phi_p}$ transforms pure quantum state~$\lvert \psi \rangle =f(\hat a^ \dagger _{\wpf}) \lvert 0 \rangle$, as follows        
        \begin{equation}
          \begin{split}
            \lvert \Psi \rangle &= \hat{\text{FF}}_{\phi_p} \lvert \psi \rangle\\
            &= \hat{\text{FF}}_{\phi_p} f(\hat a^ \dagger _{\wpf}) \lvert 0 \rangle\\
            &=  f(\hat{\text{FF}}_{\phi_p} \hat a^ \dagger _{\wpf}\hat{\text{FF}}^{\dagger}_{\phi_p}) \lvert 0 \rangle\\   
            &=  f(\hat a^ \dagger _{ \wpfo}) \lvert 0 \rangle\, ,    
        \end{split}
        \label{eqn:ffpsi}
      \end{equation}
      where the output photon-wavefront~$\wpfo$ is given by Eq.~\eqref{eqn:wpfii'(x,y)}.
      Therefore, similar to all linear optical operators, a quantum 4f-processing system without changing the photon statistics and the quantum light representation in the Fock space transforms only the associated single-photon wavefunction (its wavefront). 
      Let us consider two practically common cases of Fock representation for the input quantum light of the 4f-processor.
      \begin{itemize}
      \item \textit{ Example 1: Single-photon input}
        \par
        The 4f-processing operator~$\hat{\text{FF}}_{\phi_p}$ transforms single-photon number state~$\lvert 1 \rangle_{\wpf}= \hat a^{\dagger}_{\wpf}\lvert 0 \rangle$ as follows
        \begin{equation}
          \begin{split}
            \hat{\text{FF}}_{\phi_p} \lvert 1 \rangle_{\wpf}&= \hat{\text{FF}}_{\phi_p}  \hat a^{\dagger}_{\wpf}\lvert 0 \rangle\\
            &=  \hat a^{\dagger}_{\wpfo}\lvert 0 \rangle 
            \\
            &=\iint dx dy   \,  \wpfo (x,y) \hat a^\dagger_{x,y}\lvert 0 \rangle             
            \\
             &= 
             \iint dx dy   \,   \wpf (-x,-y)\ast h_{\hsub}(x,y) \hat a^\dagger_{x,y}\lvert 0 \rangle              \\
             &=
             \iint dx dy   \,   \wpf (-x,-y)\ast h_{\hsub}(x,y) \lvert 1 \rangle_{x,y}    \, .                          
            \end{split}
          \end{equation}
          The above equation shows that a quantum 4f-processing system performs a single-photon quantum state convolution with the impulse response function of the system, a quantum operation that can bring many applications for quantum technology.
                \item \textit{ Example 2: Glauber State input}
        \par
        The output of 4f-processor for Glauber state input~
        $\lvert\alpha \rangle_{\wpf}=\exp( -|\alpha|^ 2/2 + \alpha \hat a^ \dagger _{\wpf})\lvert 0 \rangle$,
        is Glauber state~ $\hat{\text{FF}}_{\phi_p}\lvert\alpha \rangle_{\wpf}=\exp( -|\alpha|^ 2/2 + \alpha \hat a^ \dagger _{\wpfo})\lvert 0 \rangle=\lvert\alpha \rangle_{\wpfo}$,
        where~$\wpfo$ is given by Eq.~\eqref{eqn:wpfii'(x,y)}.
        \end{itemize}
        \subsection{Quantum 4f-Processing System With Periodic Pupil}
        Consider a quantum 4f-processing system with a periodic pupil phase factor; in other words, the pupil is a two-dimensional grating-like spatial phase modulator discussed in section~\ref{sec:grating}.
        Assume its pupil phase factor~$\Pf(x,y)= e^{-i \phi_p(x,y)}$ has periods~$\xg$ and $\yg$ in the $x$ and $y$ directions, respectively.
        Therefore, the corresponding spatial angular frequencies are $\kappa_x=\frac{2 \pi }{\xg}$ and $\kappa_y=\frac{2 \pi }{\yg}$, and the Fourier expansion of the pupil phase factor~ $\Pf(x,y)$ is as follows
        \begin{equation}
          \begin{split}
            \Pf(x,y) &=\sum_r\sum_s \prs_{rs} e^{i\left( r\kappa_x x + s\kappa_y y\right)} \, ,\\
          \end{split}
          \label{eqn:ft[ffp]main}
        \end{equation}
        where $\prs_{rs}$ is the two-dimensional Fourier coefficient of the pupil phase factor (see appendix~\ref{sec:pppf}).
        \par
        Appendix~\ref{sec:pppf} shows that the impulse response of such a 4f-processor becomes the discrete lattice-like function~\eqref{eqn:hffdiscapp}:        
        \begin{equation}
          \begin{split}
            h_{\hsub}(x,y)
            &= - 2 \pi  e^{i 4 \mko f}   \sum_r\sum_s \prs_{rs}  \delta(x- \lhx{r}) \delta( y- \lhy{s}) \, ,               
          \end{split}
          \label{eqn:hffdiscgen}
        \end{equation}
        where $( \lhx{r}, \lhy{s})= (r\lhx{1},s\lhy{1})$ is associated with lattice-point~$(r,s)$ and the 4f-impulse response lattice-constants $\lhx{1}$ and $\lhy{1}$ are defined as follows
        \begin{equation}
          \begin{split}
          \lhx{1}&=\frac{f \kappa_x}{\mko}=\frac{f c \kappa_x}{\moo}\\
          \lhy{1}&=\frac{f \kappa_y}{\mko}=\frac{f c \kappa_y}{\moo}\, ,
        \end{split}
        \label{eqn:lxy'}
      \end{equation}
      where $f$ denotes the confocal length of the 4f-processor's lenses.      
      The impulse response equation~\eqref{eqn:hffdiscgen} simplifies the 4f-transformation~\eqref{eqn:ffadff} as follows (see Eq.~\eqref{eqn:adffoutperiodic})
      \begin{equation}
          \begin{split}
            \hat{\text{FF}}_{\phi_p}\, \hat a^ \dagger _{\wpf}\, \hat{\text{FF}}^{\dagger}_{\phi_p}&=\hat a^ \dagger _{\wpfo}\\
          &=  -e^{i 4 \mko f}\sum_{r,s}  \prs_{rs} \,  \hat a^\dagger_{\bar{\wpf}_{r,s}}\, ,
            \end{split}
          \label{eqn:ffadffperiodic}
        \end{equation}
        where wavefunction~$\bar{\wpf}_{r,s}$ is the parity transformed and displaced by~$(\lhx{r}, \lhy{s})$ of wavefunction~$\wpf$:
        \begin{equation}
          \bar{\wpf}_{r,s} (x,y)= {\wpf} (\lhx{r}-x,\lhy{s}-y)\, .
          \end{equation}
        One may drop the trivial constant phase factor~$ -e^{i 4 \mko f}$ in Eq.~\eqref{eqn:ffadffperiodic}, which is due to the net displacement in the 4f-system, but for the sake of completeness, let us keep the factor.
        Assume $(\cen{x},\cen{y})$ and $(\Delta x, \Delta y)$ denote the center and the width of the photon-wavepacket~$\wpf$, respectively, such that $\wpf(x,y)\approx 0 \ \forall \ x \notin \left(\cen{x}-\frac{\Delta x}{2}, \cen x +\frac{\Delta x}{2}\right)\,\cup\,y \notin \left(\cen{y}-\frac{\Delta y}{2}, \cen y + \frac{\Delta y}{2}\right)$.
        Furthermore, assume photon-widths are narrower than their corresponding 4f-impulse response lattice-constants, $\Delta x < \lhx{1}$ and $\Delta y <\lhy{1}$.
        In this case, the wavepackets~$\bar{\wpf}_{r,s}$ with a central point at~$\left(\lhx{r}-\cen{x},\lhy{s}-\cen{y}\right)$ are orthogonal to each other, $\langle \bar{\wpf}_{r,s},\bar{\wpf}_{r',s'}\rangle=\delta_{r,r'}\delta_{s,s'}$, and therefore, the corresponding creation operators hold the canonical commutation relation (see also Eq.~\eqref{eqn:[axi,adxi]}) as follows        
        \begin{equation}
          \begin{split}
            \left[ \hat a _{\bar{\wpf}_{r,s}}, \hat a^ \dagger _{\bar{\wpf}_{r',s'}}\right]           & =\langle \bar{\wpf}_{r,s} \vert \bar{\wpf}_{r',s'} \rangle=\delta_{r,r'}\delta_{s,s'}\, .
            \end{split}
          \label{eqn:[axi,adxi]}
        \end{equation}
        Since there is no overlap between wavepackets of different lattice points, they are orthogonal.
        To clarify the above 4f-transformation, let us write the photon-wavepacket creation operators simply with a subscript of their central points.
        Then, we substitute the input creation operator~$\hat a^ \dagger _{\wpf}$ with $\hat a^ \dagger _{\cen{x},\cen{y}}$, and the parity transformed displaced output creation operators~$  \hat a^\dagger_{\bar{\wpf}_{r,s}}$ with $\hat b^\dagger_{\lhx{r}-\cen{x}, \lhy{s}-\cen{y}}$.
        Therefore Eq.~\eqref{eqn:ffadffperiodic} takes the form as follows
        \begin{equation}
          \begin{split}
            \hat{\text{FF}}_{\phi_p}\, \hat a^ \dagger _{\cen{x},\cen{y}}\, \hat{\text{FF}}^{\dagger}_{\phi_p}
          &=  -e^{i 4 \mko f}\sum_{r,s}  \prs_{rs} \,  \hat b^\dagger_{ \lhx{r}-\cen{x}, \lhy{s} - \cen{y}}\, .
            \end{split}
          \label{eqn:ffadxyffperiodic}
        \end{equation}
        Equation~\eqref{eqn:ffadxyffperiodic} denotes that a quantum 4f-processing system with a periodic, grating-like pupil maps a single-photon with a wavefront localized at point~$\left(\cen{x},\cen{y}\right)$ on the input plane into a lattice-like wavefront with lattice coordinates~$\left(\lhx{r}-\cen{x}, \lhy{s}-\cen{y}\right),\ r,s=0, \pm 1, \pm2, \hdots$, on the output plane.
        Also, the wavefront's probability amplitude at the lattice-point~$\left( \lhx{r}-\cen{x}, \lhy{s}-\cen{y}\right)$ corresponds to~$\prs_{rs}$.
        \section{Quantum Pulse Shaping via an 8\lowercase{f}-Processing System}
        \begin{figure*}[!t]
          \centering
          \includegraphics[width=\textwidth]{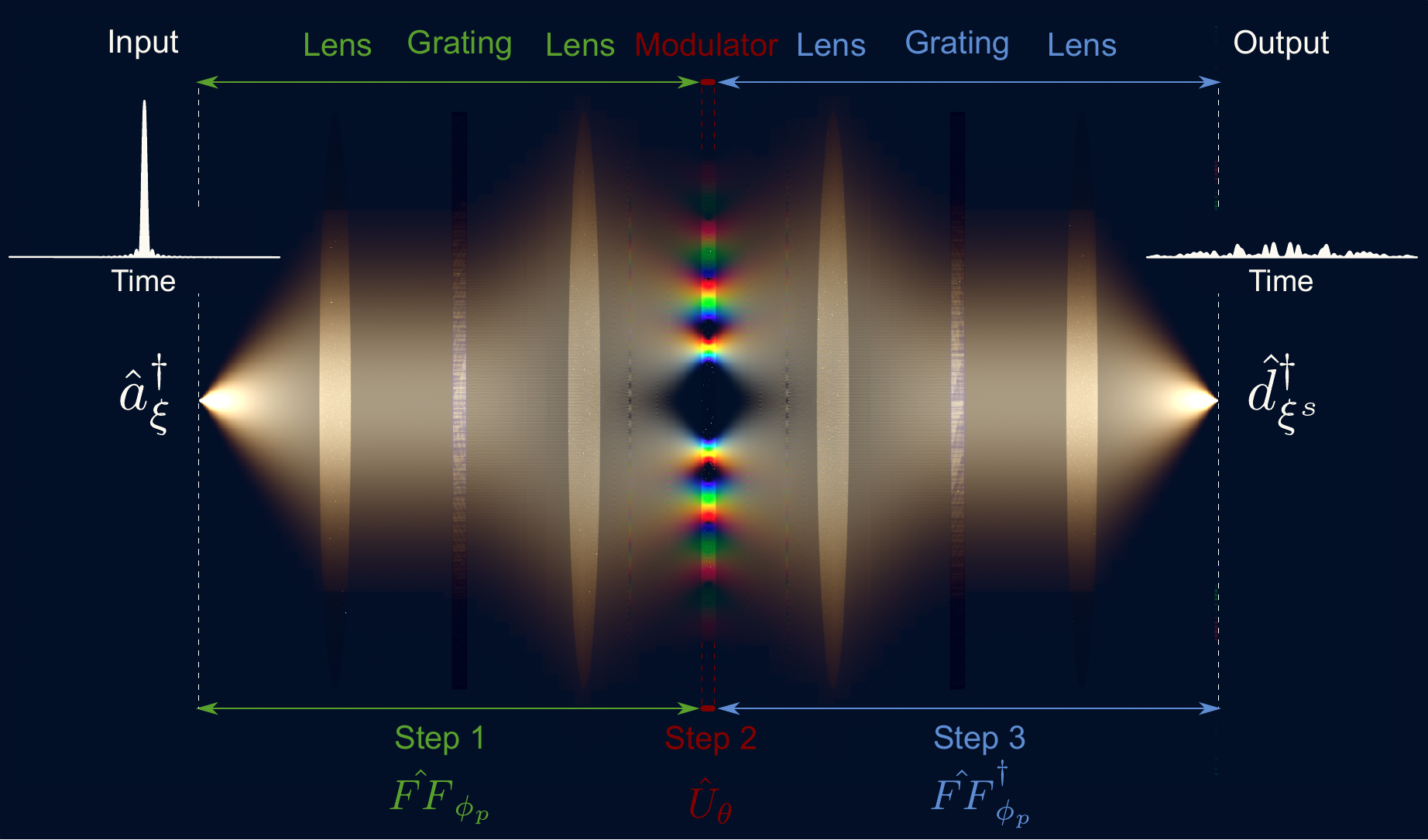}
          \caption{\textbf{Quantum Pulse Shaping.}
            A quantum pulse with the temporal shape shown at the upper-left corner enters the pulse shaper.
            First, the 4f-processor~$\hat{\text{FF}}_{\phi_p}$ in step 1 decomposes the frequency component of the input quantum pulse.
            Then, at step 2, each spatially separated frequency component~$\omega$ of the input pulse gets the desired phase via the spatial phase modulator~$\hat{\text{U}}_{\theta}$.
            Eventually, at step 3, the inverse of the first unitary 4f-operation, i.e., $\hat{\text{FF}}^{-1}_{\phi_p}=\hat{\text{FF}}^{\dagger}_{\phi_p}$, reassembles the spatially separated frequency components into a single spatial mode.
            Consequently, the input photon-wave packet’s temporal shape, shown in the upper-left corner, gets reshaped at the output, shown in the upper-right corner. 
            The simulation considers a rectangular function as the spectral photon-wavepacket (the sinc function as the temporal photon-wavepacket) for the input of the pulse shaper.
            The reshaping of the pulse is due to 31 random phase shifts applied to 31 spectral chips.
            The simulation uses a wavelength-based color scheme to propagate each photon-wave packet's frequency component through the pulse shaper.
          }\label{fig:ps}
        \end{figure*}
        Optical pulse shaping techniques have many applications in classical and quantum technologies such as optical and quantum communications~\cite{salehi_jlwt_1990, rezai_ieeeit_2021} and  quantum computations~\cite{kues_np_2019}.
       Due to its importance, in this section, we utilize the aforementioned mathematical tools to detail a quantum pulse shaping procedure.
        \par        
        A quantum pulse shaper phase modulates different frequency components of quantum light by any desired phases from $0$ to $2\pi$~\cite{rezai_ieeeit_2021}.
        Therefore, a quantum pulse shaper is, in fact, a quantum spectral phase modulator.
        Hence, in this section of the paper, we exceptionally assume photons occupy multiple or continuous spectral modes (see appendix~\ref{sec:mfm}).
        \par
        Spectral phase modulation or pulse shaping can be realized in three steps, as Fig.~\ref{fig:ps} shows.
        The first step spatially separates the frequency components of the photon. 
        In the second step, a spatial phase modulation phase shifts each frequency component of the photon with the desired or preassigned phase values.
        Finally, The third step brings the spatially separated frequency phase modulated components into a single spatial mode.                
        \subsection{Step one: frequency separation via a 4f-processor}
        \label{sec:ps_step1}
        For the sake of simplicity, this section assumes a one-dimensional photon-wavefront in the $x$-direction, and the light propagation is in the $z$-direction.
        Consider a single photon with a narrow, Airy disk-like, diffraction-limited wavefront localized at the center of the input plane of a 4f-system, $\cen{x}=0$.
        As Eq.~\eqref{eqn:axioapp} shows, since the features of this narrow wavefront would have minimal consequence, we write its corresponding creation operator as follows
         \begin{equation}
          \begin{split}
            \hat a^ \dagger _{\wpf} 
            &=\int d\omega   \,  \wpf (\omega) \hat a_{0}^\dagger(\omega)\, ,       
            \end{split}
          \label{eqn:axio0}
        \end{equation}
        where $\hat a^\dagger_{0}\equiv\hat a^\dagger_{\cen{x}=0}$, and $\wpf(\omega)$ denotes the photon's spectral wavepacket or probability amplitude.
        Furthermore, take a periodic, phase-only, grating-like pupil for the 4f-system.
        Therefore, the pupil Fourier transform~\eqref{eqn:ft[ffp]main} becomes: (see also the corresponding diffraction grating equation~\eqref{eqn:ft[ephig]})
        \begin{equation}
          \begin{split}
            \Pf(x) &=e^{-i \phi_p(x)}=\sum^{R}_{r=-R} \prs_{r}e^{i\left( r\kappa_x x\right)} \, ,\\
          \end{split}
          \label{eqn:ft[ffp]mainx}
        \end{equation}
        where $r=\pm R$ corresponds to the maximum spatial frequency of the pupil phase factor in the $x$-direction, i.e., $\prs_{r}\approx 0 \ \forall \ \abs{r}>R$.
        We obtained the quantum 4f-transformation~\eqref{eqn:ffadxyffperiodic} under the single frequency mode assumption.
        Technically, the chromatic aberration of its element (especially the lenses) can be corrected.
        Therefore, such a chromatic aberration corrected quantum 4f-processor transforms each frequency component~$\hat a_{0}^\dagger(\omega)$ of the creation operator equation~\eqref{eqn:axio0}  according to Eq.~\eqref{eqn:ffadxyffperiodic}, which gives
        \begin{equation}
          \begin{split}
            \hat{\text{FF}}_{\phi_p}\, \hat a_{\rpx}^ \dagger  (\omega)\, \hat{\text{FF}}^{\dagger}_{\phi_p}
           & =   -e^{i 4 \mko f}  \sum^{R}_{r=-R} \prs_{r} \, \hat b^\dagger_{\mrpx \lpx{r}} (\omega)            
     \, ,
            \end{split}
          \label{eqn:ffadxffperiodic1d} 
        \end{equation}
        where $\lpx{r}=\mrpx r \lpx{1}$ indicates the $r$th lattice coordinate of the frequency component~$\omega$ and is equivalent to its $r$th order of diffraction.
    We have added frequency~$\omega$ as a subscript to the lattice-constant and lattice coordinates to highlight their frequency dependency.   
        This section considers spectrally multi-mode (continuous mode) photons, and equation~\eqref{eqn:lxy'}, $\lpx{1}=\frac{ f\, c\, \kappa_x}{\omega}$, denotes that the lattice-constant~$\lpx{1}$ of the output photon-wavefront depends on the photon's angular frequency~$\omega$.
        Furthermore, angular frequency~$\omega$ is given as an argument to the creation operators~$\hat a_{\rpx}^ \dagger(\omega)$ and~$\hat b^\dagger_{\mrpx r \lpx{1}} (\omega)$ to emphasize that Eq.~\eqref{eqn:ffadxffperiodic1d} is a single frequency mode transformation.
        Finally, equation~\eqref{eqn:ffadxffperiodic1d} indicates that the spatial range of the discretized photon-wavefront at the output of the 4f-processor is $-\lpx{R}\leq x \mrpx \leq \lpx{R}$.
        \par
        Let~$\hat b^\dagger$ denote the one-dimensional unitary 4f-operation~\eqref{eqn:ffadxffperiodic1d} on the frequency-dependent photon-wavepacket creation operator~\eqref{eqn:axioapp}:      
        \begin{equation}
          \begin{split}
             \hat b^\dagger&=\hat{\text{FF}}_{\phi_p}\, \hat a^ \dagger _{ \wpf} \, \hat{\text{FF}}^{\dagger}_{\phi_p} \\
             &= \int d\omega   \,  \wpf (\omega) \, \hat{\text{FF}}_{\phi_p}\, \hat a_{\rpx}^ \dagger(\omega) \, \hat{\text{FF}}^{\dagger}_{\phi_p}\\
             &=  -\sum^{R}_{r=-R} \prs_{r} \int d\omega   \,  \wpf (\omega) \, e^{i \omega \frac{4 f}{c}} \, \hat b^\dagger_{ \lpx{r}}(\omega)
             \, .
            \end{split}
          \label{eqn:ffaxioffd}
        \end{equation}
        The frequency dependency of the output lattice-constant~$ \lpx{r}$ implies that the 4f quantum operator~$\hat{\text{FF}}_{\phi_p}$ spatially separates frequency components.
        It separates different frequency components of the spatially single-mode and spectrally multimode input photon creation operator~$\hat a^ \dagger _{\wpf}$ into different lattices with different lattice-constant~$\lpx{r}$.       
        Yet, a lattice-point of one frequency component of the input photon may overlap with a lattice-point of another frequency component.
        For example, the central lattice-point~$\lpx{r}=r\lpx{1}=0$ for $r=0$ is the common lattice-point for all frequencies, and $p_{r=0}$ is the probability amplitude that the input photon emerges at the center.
        However, for a perfect quantum pulse shaping, we need to completely separate different frequency components of the input photons.
        As discussed in appendix~\ref{app:psuc}, the conditions for ideal frequency separation via a 4f quantum processor are that the Fourier transform of the pupil phase factor~\eqref{eqn:ft[ffp]mainx} has no DC term ($p_{r=0}$).
        And the photon bandwidth~$\Delta \omega$ is shorter than $\frac{\omega_{max}}{R}$, where $\omega_{max}$ is the maximum of the angular frequency components that the input photons occupy, and $R$ corresponds to the maximum spatial frequency of the phase-only grating (pupil).        
        Therefore, it is possible to find the frequency  mapped function~$\Omega(x)$, determining the frequency component mapped to position~$x$, which, as shown in appendix~\ref{sec:fmf}--Eq.~\eqref{eqn:Omega(x)app}, is  
        \begin{equation}
           \begin{split}
             \Omega(x)
            &=  \min(R,\floor{\frac{\abs{x}}{\lpxomax{1}}})\frac{\lpxomax{1}}{\abs{x}} \omega_{max}\, ,
           \end{split}
           \label{eqn:Omega(x)}
         \end{equation}
         where $\floor{x}$ and $\abs{x}$ denote the floor function of $x$ and the absolute value function of $x$, respectively.
         This means function $\Omega(x)$ on the $r$th diffraction order position~$\lpx{r}=rx^{(1)}(\omega)$ of the frequency~$\omega$  gives the corresponding frequency, i.e., $ \Omega(\lpx{r}) =\omega,\ \forall \ 1 \leq r \leq R, \   \omega_{max}-\Delta \omega<\omega < \omega_{max}$, see appendix~\ref{sec:fmf}--Eq.~\eqref{eqn:Omega(x(omega0))} for details.
        \subsection{Step two: frequency modulation via a spatial phase modulator }
        As step one spatially separates the frequency components, now for the second step, a spatial phase modulator~\ref{sec:spmo}, as is discussed in this section, can apply any desired phase into any selected spectral element.
        Thus, for example, assume $\theta(\omega)$ is the desired spectral phase modulation and name its corresponding one-dimensional spatial phase modulator as~$\hat{\text{U}}_{\theta}$, which, as Eq.~\eqref{eqn:Uphixy} denotes, can be written as
        \begin{equation}
          \hat{\text{U}}_{\theta}  = e^{- i \sum_{x} \phi(x) \hat n_{x}} \, .
          \label{eqn:Uphix}
        \end{equation}
        The amount of phase modulation by a practical spatial phase modulator usually depends on the frequency of the incident light beam.
        But spatial phase modulator equation~\eqref{eqn:Uphixy} (also Eq.~\eqref{eqn:Uphix}) is valid either for an ideal frequency-independent spatial phase modulator or at least when the incident light beam on each point~$x$ of the modulator is a single spectral mode, which is the case in this section.   
        Because on position~$x$ at the output plane of the first 4f-processor, only a single frequency ($\omega=\Omega(x)$) component may be mapped.
        Therefore, for the reduced Hilbert space of the single spatial mode input~\eqref{eqn:axio0}, Eq.~\eqref{eqn:Uphix} is simplified, such that $\hat n_{x}\equiv \hat b_{x}^{\dagger} (\Omega(x)) \hat b_{x} (\Omega(x))=\hat n_{x} (\Omega(x)) $, and $ \phi(x)=\theta(\Omega(x))$, where~$\Omega(x)$ is given by Eq.~\eqref{eqn:Omega(x)}.
        Consequently, Eq.~\eqref{eqn:Uphix} takes the form as
        \begin{equation}
          \hat{\text{U}}_{\theta}  = \prod_{x}e^{- i  \theta(\phi(\omega(x))\ \hat n_{x} (\Omega(x)) } \, .
          \label{eqn:Uthetax}
        \end{equation}
        If the distance~$x-\lpx{r}$ is  more significant than the size of the point spread function, the number operator at point $x$ ($\hat n_{x}  (\Omega(x))$) commutes with the creation operator at point~$\lpx{r}$ ($\hat b^\dagger_{\lpx{r}}(\omega)$).
        Therefore, the operation of the spatial phase modulator~$\hat{\text{U}}_{\theta}$ on the creation operator~$\hat b^\dagger_{\lpx{r}}(\omega)$ can be simplified as follows       
        \begin{equation}
          \begin{split}
            &\hat{\text{U}}_{\theta} \, \hat b^\dagger_{\lpx{r}}(\omega) \hat{\text{U}}^{\dagger}_{\theta}\\
            &=e^{- i \theta(\Omega(\lpx{r})) \hat n_{\lpx{r}} (\Omega(\lpx{r}))} \hat b^\dagger_{\lpx{r}}(\omega) e^{- i \theta(\Omega(\lpx{r})) \hat n_{\lpx{r}}  (\Omega(\lpx{r}))}\\
            &=e^{- i \theta(\omega) \hat n_{\lpx{r}} (\omega)} \hat b^\dagger_{\lpx{r}}(\omega) e^{- i \theta(\omega) \hat n_{\lpx{r}}  (\omega)}\\
            &=e^{- i \theta(\omega) } \hat b^\dagger_{\lpx{r}}(\omega) \, ,
             \end{split}
          \label{eqn:ubdupsst}
        \end{equation}
        where Eq.~\eqref{eqn:Omega(x(omega0))}, $ \Omega(\lpx{r}) =\omega$, is used.
        The unitary spatial phase modulation equation~\eqref{eqn:Uthetax} transforms the 4f-processor’s output~\eqref{eqn:ffaxioffd} into a  single-photon creation operator~$\hat{c}^{\dagger}$        
        \begin{equation}
          \begin{split}
            \hat c^{\dagger} &=\hat{\text{U}}_{\theta} \hat b^ \dagger \hat{\text{U}}^{\dagger}_{\theta}   \\    
            &=  -\sum^{R}_{r=-R} \prs_{r} \int d\omega   \,  \wpf (\omega) \, e^{i \omega \frac{4 f}{c}} \, \hat{\text{U}}_{\theta}  \hat b^\dagger_{ \lpx{r}}(\omega) \hat{\text{U}}^{\dagger}_{\theta}             \\
            &=  -\sum^{R}_{r=-R} \prs_{r}  \int d\omega   \,  \wpf (\omega) \, e^{i \omega \frac{4 f}{c}}  e^{- i \theta(\omega) } \hat b^\dagger_{\lpx{r}}(\omega) \, ,
            \end{split}
          \label{eqn:uthetaffaxio}
        \end{equation}
        where Eq.~\eqref{eqn:ubdupsst} is used.
        Therefore, Eq~\eqref{eqn:uthetaffaxio}, the output of step two, implies that the spatial modulator~\eqref{eqn:Uthetax} adds frequency-dependent phase factor~$e^{- i \theta(\omega) }$ to the corresponding spectral component of the input operator~\eqref{eqn:ffaxioffd}.
        \subsection{Step three: beam assembly via a 4f-processor }
        The purpose of the third step is to gather the spatially separated and appropriately (spectrally) phase-modulated components of the creation operator~\eqref{eqn:uthetaffaxio}, the output of the second step, into a single spatial mode.
        Reassembling procedure can be realized by the inverse operation of the first step's 4f-operator, $\hat{\text{FF}}^{-1}_{\phi_p}=\hat{\text{FF}}^{\dagger}_{\phi_p}$.
        In other words, in step three, the 4f-processor~$\hat{\text{FF}}^{\dagger}_{\phi_p}$, whose pupil phase factor~$\iPf(x)$ is related to the first 4f-processor~$\Pf(x)$ as $\iPf(x)=\Pf^{\ast}(-x) =e^{i \phi_p(-x)}$ (see appendix~\ref{app:i4f}-Eq.~\eqref{eqn:p-p'}), can reassemble the single-photon creation operator~$ \hat c^{\dagger}$ (Eq.~\eqref{eqn:uthetaffaxio}) into a single spatial mode.
        To demonstrate that, using Eq.~\eqref{eqn:ffadxyffperiodic}, we first apply the one-dimensional 4f-transformation~$\hat{\text{FF}}^{\dagger}_{\phi_p}$ on operator~$\hat b^\dagger_{\lpx{r}}(\omega)$:
        \begin{equation}
          \begin{split}
            \hat{\text{FF}}^{\dagger}_{\phi_p}\, \hat b^ \dagger _{\lpx{r}} (\omega)\, \hat{\text{FF}}_{\phi_p}
            &=  -e^{i 4 \mko f}\sum^{R}_{r'=-R}  \iprs_{r'} \,  \hat d^\dagger_{ \lpx{r'}- \lpx{r}} (\omega)\\
            &=  -e^{i 4 \mko f}\sum^{R}_{r'=-R}  \prs^{\ast}_{r'} \,  \hat d^\dagger_{ \lpx{r'-r}} (\omega)     \, ,
            \end{split}
          \label{eqn:ffbdxffperiodic1d(x)} 
        \end{equation}
        where the single-photon creation operator $\hat d^\dagger_{ \lpx{r'-r}} (\omega)$ is associated with
        the output plane of the 4f-processor~$\hat{\text{FF}}^{-1}_{\phi_p}=\hat{\text{FF}}^{\dagger}_{\phi_p}$.
        And also, Eq.~\eqref{eqn:prs-prs'}, $\iprs_{r'}=\prs^{\ast}_{r'} $, is used, where $\prs_{r'}$ is the corresponding Fourier coefficient of 4f-processor~$\hat{\text{FF}}_{\phi_p}$.
        Using Eq.~\eqref{eqn:ffbdxffperiodic1d(x)}, the 4f-processor~$\hat{\text{FF}}^{\dagger}_{\phi_p}$ transforms the input~\eqref{eqn:uthetaffaxio} as follows
        \begin{equation}
          \begin{split}
            \hat{\text{FF}}^{\dagger}_{\phi_p} \hat c^{\dagger}  \hat{\text{FF}}_{\phi_p}        
               &=  \sum^{}_{r,r'} \prs_{r} \prs^{\ast}_{r'} 
               \int d\omega  \wpf^{s} (\omega)\,   \hat d^\dagger_{\lpx{r'-r}} (\omega)\, ,
             \end{split}
          \label{eqn:upspre}
        \end{equation}
        where
        \begin{equation}
          \begin{split}
             \wpf^{s} (\omega)&=\wpf (\omega) \, e^{i \omega \frac{8 f}{c}}  e^{- i \theta(\omega) }\, .
            \end{split}
          \label{eqn:wps}
        \end{equation}
        Change of variables $r,r'$ to $r,r''$, as $r''=r'-r$, allows us to sum over the index~$r$, which gives $\sum^{}_{r} \prs_{r} \prs^{\ast}_{(r+r'')}=\delta_{r'',0}$, where Eq.~\eqref{eqn:unitaryp} is used.
        Therefore, position~$\lpx{r'-r}=\lpx{r''}=r''\lpx{1}$, due to $\delta_{r'',0}$, becomes frequency-independent and zero, $\lpx{r''}=\lpx{0}=0$, and thus reduces Eq.~\eqref{eqn:upspre} into a single spatial mode located at the central lattice-point~$\lpx{0}=0$:         
        \begin{equation}
          \begin{split}
         \hat{\text{FF}}^{\dagger}_{\phi_p} \hat c^{\dagger}  \hat{\text{FF}}_{\phi_p}          
               &= 
               \int d\omega  \wpf^{s} (\omega)\,   \hat d^\dagger_{0} (\omega)\\\
               &=\hat d^\dagger_{\wpf^{s}} \, .
             \end{split}
          \label{eqn:upspre0}
        \end{equation}
        Equation~\eqref{eqn:upspre0} gives the output of the pulse shaper.
        Therefore, combining the above three steps quantum transformation equations~\eqref{eqn:ffaxioffd}, \eqref{eqn:uthetaffaxio} and \eqref{eqn:upspre0} make the quantum pulse shaper transformation as
        \begin{equation}
          \begin{split}
             \hat{\text{PS}}\, \hat a^ \dagger _{ \wpf}\, \hat{\text{PS}}^{\dagger} 
               &=\hat d^\dagger_{ \wpf^{s}} \, .
             \end{split}
          \label{eqn:ups}
        \end{equation}        
        where
        \begin{equation}
          \hat{\text{PS}}=\hat{\text{FF}}^{\dagger}_{\phi_p} \hat{\text{U}}_{\theta}  \hat{\text{FF}}_{\phi_p}
        \end{equation}
        denotes the quantum pulse-shaping operator.
        Equation~\eqref{eqn:ups} indicates that the pulse shaper's output creation operator~$\hat d^ \dagger _{ \wpf^{s}}$ becomes spatially single-mode at position $x=0$, similar to the input creation operator~$\hat a^ \dagger _{ \wpf}$.
        However, the output spectral photon-wavepacket is transformed from~$\wpf$ to $\wpf^{s}$, given by Eq.~\eqref{eqn:wps}, which eventually reshapes the pulse or the probability of the photon-wavepacket in the time domain~\cite{rezai_ieeeit_2021}, see Fig.~\ref{fig:ps}.
        It is worth noting that the phase factor~$ e^{i \omega \frac{8 f}{c}}$, in the shaped wavepacket $\wpf^{s} (\omega)$, is due to the net displacement transition (section~\ref{sec:dto}) by $z=8f$ through the 8f-pulse shaper.
        This phase factor is equivalent to a time evolution (see section~\ref{sec:teo}) by $t=-\frac{8f}{c}$. 
        Thus, ignoring the trivial phase factor~$ e^{i \omega \frac{8 f}{c}}$, the quantum pulse shaper~$\hat{\text{PS}}$ has shaped the photon-wavepacket~$\wpf(\omega)$ into $\wpf^{s} (\omega)=\wpf (\omega) e^{- i \theta(\omega) }$, meaning the photon component of frequency~$\omega$ got its corresponding phase factor~$e^{- i \theta(\omega) }$.
        \subsection{Examples}
        As before, we take the single-photon state and the Glauber state as the inputs to the quantum pulse shaper to clarify its operation!
        \begin{itemize}
        \item \textit{ Example 1: Single-photon input}          
          \par
          Consider a single-photon number state with spectral-wavepacket~ $\wpf(\omega)$ (see Eq.~\eqref{eqn:axio0}), which corresponds to the quantum state          
         \begin{equation}
          \begin{split}
            \lvert 1 \rangle_{\wpf}= \hat a^ \dagger _{\wpf}  \lvert 0 \rangle
            &=\int d\omega   \,  \wpf (\omega)   \lvert 1\rangle_{\omega }\\
            &\equiv\int d\omega   \,  \wpf (\omega)   \lvert \omega \rangle\, , 
            \end{split}
          \label{eqn:ps_sph_input}
        \end{equation}
        as the input of the pulse shaper.
        State~$\lvert 1 \rangle_{ \omega}$ denotes continuous mode single-photon number state with angular frequency~$\omega$.
        The pulse shaping operation~\eqref{eqn:ups} transforms the single-photon quantum state~\eqref{eqn:ps_sph_input} as follows
        \begin{equation}
          \begin{split}
            \hat{\text{PS}} \lvert 1 \rangle_{\wpf}&= \hat{\text{PS}} \hat a^{\dagger}_{\wpf}\lvert 0 \rangle =\hat{\text{PS}}\hat a^{\dagger}_{\wpf}\hat{\text{PS}}^{\dagger}\lvert 0 \rangle=\hat d^{\dagger}_{\wpfi^{s}}\lvert 0 \rangle \\
            &=\lvert 1 \rangle_{\wpfi^{s}} =  \int d\omega   \,   e^{i \omega \frac{8 f}{c}} e^{- i \theta(\omega) } \wpf (\omega)   \lvert \omega \rangle\, ,
             \end{split}
          \label{eqn:ps_sph}
        \end{equation}        
        where Eq.~\eqref{eqn:wps} is used.
        The applied frequency-dependent phase~$\theta(\omega)$ would reshape the input single photon's quantum state in the time domain, hence the name~\cite{kues_np_2019, rezai_ieeeit_2021} (see Fig.~\ref{fig:ps}).
        \item \textit{ Example 2: Glauber State input}
          \par 
          For a Glauber state input~$\lvert\alpha \rangle_{\wpf}$,
        the pulse shaper gives another Glauber state with the same amplitude~$\alpha$:
        \begin{equation}
          \begin{split}
            \hat{\text{PS}} \lvert \alpha \rangle_{\wpf}&=\hat{\text{PS}}e^{-|\alpha|^ 2/2 + \alpha \hat a^ \dagger _{\wpf}} \hat{\text{PS}}^{\dagger}\lvert 0 \rangle\\
            &=e^{ -|\alpha|^ 2/2 + \alpha \hat d^ \dagger _{\wpfi^{s}}}\lvert 0 \rangle\\
            &=\lvert\alpha \rangle_{\wpfi^{s}}\, .
          \end{split}
        \end{equation}
        Yet, each photons' wavepacket is transformed from $\wpf(\omega)$ to function $\wpfi^{s}(\omega)$.
        \end{itemize}
        
\section{Conclusion}
        The complete representation of the quantum state of light boils down to two aspects.
        One is the quantum light's photon statistics or its Fock representation (represented by function~$f(\hat a^{\dagger})$ in this paper).
        For example, coherent, single-photon, and squeezed states of light are discriminated against due to their different representation in the Fock space.
        The other is the quantum wave function or wavepacket (usually symbolized by $\xi$, and appears as $\hat a^{\dagger}_\xi$ in this paper) of each constituent photon of the quantum light.
        The spatial mode (image) or the wavefront of photons is one essential feature of the photon wavepacket ($\xi$) that brings vast information capacity. 
        \par
        Developing the quantum model of optical components is essential for various quantum information processing.
        For example, the quantum model of a simple beam splitter~\cite{leonhardt_rpp_2003}, array waveguide gratings \cite{capmany_oe_2013}, starcoupler and spectral phase shifter~\cite{rezai_ieeeit_2021} have a significant impact on the mathematical modeling of quantum operations and applications.
        This paper mathematically models Fourier optical systems' components that transform photons' wavefronts and, consequently, process the information encoded on them.
        \par
        It introduces the fundamental concepts of quantum Fourier optics and shows that unitary quantum Fourier optics operations can be mathematically modeled with a sequence of phase-shifting operations.
        Accordingly, it analyses the main building blocks of quantum Fourier optics, namely lens operator and 4f-processors, and shows that lens operators and 4f-processors perform continuous mode quantum Fourier transformation and convolution on the quantum state of a single photon.
        \par
        Furthermore, this paper details the quantum pulse shaping procedure as an actual application of the concepts mentioned above. This technique plays a crucial role in quantum information processing and computation based on frequency encoded optical combes~\cite{lukens_o_2017,kues_np_2019} and is employed in quantum code division multiple access communication systems~\cite{rezai_ieeeit_2021}.
        The concepts and mathematical models introduced in this paper would benefit quantum technologies extensively. Its application ranges from quantum computation and communications and quantum orbital angular momentum information processing to quantum sensing, quantum imaging, quantum radar and quantum multiple-input/multiple-output antennas.
\appendix
   \section{Photon-Wavepacket Creation Operator}
   \label{sec:csmco}
   \begin{figure}[!t]
     \centering
          \includegraphics[width=\columnwidth]{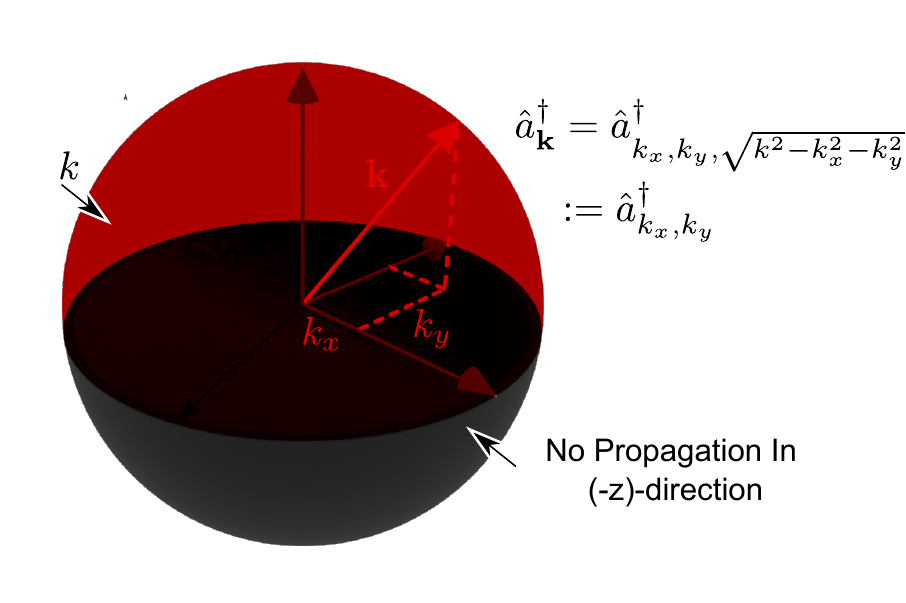}
          \caption{\textbf{The single-frequency-mode assumption on photon wavevectors.}
            The single frequency mode assumption limits single-photons to be only in the superposition of states with wavevectors shown by the sphere.
            The no-propagation in ($-z$)-direction assumption reduces the allowed state to the upper blue hemisphere, consequently makes~$\mkx$ and~$\mky$ the sufficient parameters to locate the involved wavevectors.
          }\label{fig:sfa}
        \end{figure}
        The corresponding creation operator of a photon concerning its occupation mode can be in the most general case expanded as~\cite{scully_zubairy_1997,loudon_2000}
        \begin{equation}
          \begin{split}
            \hat a^ \dagger _{\wpf}
            &=\sum_{P=h ,v}\int  d^3K \,  \wpfi_{P}  (\mathbf{K}) \hat a ^\dagger_{P,\mathbf{K}}\, ,
            \end{split}
          \label{eqn:achikp}
        \end{equation}
        where $\hat a^\dagger_{P,\mathbf{K}}$ denotes the photon creation operator at the mode with wavevector~$\mathbf{K}$ and polarization~$P$, and the corresponding photon's probability amplitude of this mode is $\wpfi_{P}  (\mathbf{K})$, which is normalized and satisfies the canonical commutation relation
        \small
        \begin{equation}
          \begin{split}
            \left[ \hat a _{\wpf}, \hat a^ \dagger _{\wpf}\right]
            &=\sum_{P',P=h ,v}\int  d^3K'  d^3K  \wpfi^{\ast}_{P'}  (\mathbf{K'})  \wpfi_{P}  (\mathbf{K})  \left[ \hat a _{P',\mathbf{K'}}, \hat a^ \dagger _{P,\mathbf{K}}\right]\\
            &=\sum_{P',P=h ,v}\int  d^3K'  d^3K  \wpfi^{\ast}_{P'}  (\mathbf{K'})  \wpfi_{P}  (\mathbf{K})  \delta_{P,P'}\delta(\mathbf{K'}-\mathbf{K})\\
            &=\sum_{P=h ,v}\int  d^3K\,  |\wpfi_{P}  (\mathbf{K})|^2\\
            & =\langle \wpfi \vert \wpfi \rangle= \|\wpfi\|^2=1\, ,
            \end{split}
          \label{eqn:[axi,adxi]}
        \end{equation}       
       \normalsize
        where commutation relation~$\left[ \hat a _{P',\mathbf{K'}}, \hat a^ \dagger _{P,\mathbf{K}}\right]= \delta_{P,P'}\delta(\mathbf{K'}-\mathbf{K})$ is used, and $\langle \wpfi \vert \wpfi \rangle$ and $\|\wpfi\|$ denote the inner product of the single-photon wave function~$\wpfi$ with itself and its norm, respectively.
        \par
        Equations~\eqref{eqn:achikp} and~\eqref{eqn:[axi,adxi]}
        show that there are at most four degrees of freedom for the occupation mode of any photon, one discrete polarization mode, and three continuous modes of wavevector components~$(K_x,K_y,K_z)$.
        This paper assumes the polarization of photons remains invariant throughout all introduced processes, which reduces the occupation mode's degree of freedom to three.
        The photons that singularly occupy one of the mode's degrees of freedom are referred to as the so-called single-mode photons with respect to that singular mode; for example, spectrally single-mode photons singularly occupy one spectral frequency alone.
        In this case, we drop its corresponding variable from the notation of photon-wavepacket~$\wpfi_{P}$ and creation operator~$\hat a ^\dagger_{P,\mathbf{K}}$.
        Therefore,  in this paper, we leave the polarization mode's variable from the consideration and write photon-wavepacket~$\wpfi_{P} (\mathbf{K})$ simply as  $\wpfi (\mathbf{K})$ and creation operator~$ \hat a ^\dagger_{P,\mathbf{K}}$ as $ \hat a ^\dagger_{\mathbf{K}}$.
        \par
        Furthermore, Let us take a spectrally single-mode quantum light with frequency~$\moo$ and angular wavenumber $\mko=\frac{\moo}{c}$.
        The paper uses bold letters to represent wavevectors, and their corresponding non-bold letter indicates their wavenumber, $\left|\mathbf{\mko}\right|=\mko$.        
        Due to the single-frequency-mode assumption (see Fig.~\eqref{fig:sfa}), photon-wavepacket~$\wpfi (\mathbf{K})$ is nonzero only at wavevector with amplitude~$K=\frac{\moo}{c}=\mko$ (i.e., $\wpfi  (\mathbf{K}\neq\mathbf{\mko})=0$).        
        Therefore, the photon-wavepacket is proportional to $\wpfi (\mathbf{K}) \propto \delta(\mathbf{K}-\mathbf{\mko})$, which reduces one more from the mode's degree of freedom.
        Thus, in Eq.~\eqref{eqn:achikp}, the general volume integration over all wavevectors reduces to the surface integration over the sphere with radius~$\mko$ ($\oint_{\mko}$), i.e.,
        \begin{equation}
          \begin{split}
            \hat a^ \dagger _{\wpf}
            &=\int  d^3K \,  \wpfi  (\mathbf{K}) \hat a ^\dagger_{\mathbf{K}}\\
            &=\oint_{\mko}  d^2k \,  \wpfi  (\mathbf{\mko}) \hat a ^\dagger_{\mathbf{\mko}} \\
            &=\iint dk_x dk_y   \,  \wpfi (k_x,k_y) \hat a^\dagger_{k_x,k_y}\, ,
            \end{split}
          \label{eqn:axikk}
        \end{equation}             
        Furthermore, equation~\eqref{eqn:axikk} takes the wavevector's $\mkx$ and $\mky$ components as the independent variables for the integration, implying the wavevector's $\mkz$ component is $\mkz=\sqrt{\mko^2-\mkx^2-\mky^2}$, assumed the quantum light is not propagating in the ($-z$)-direction.
        In equation~\eqref{eqn:axikk}, the single-frequency-mode assumption reduced the photon creation operator~ $\hat a^\dagger_{\mathbf{K}}$ to~ $\hat a^\dagger_{\mathbf{k}}$.
        The further no-propagation in ($-z$)-direction assumption makes~$\mkx$ and~$\mky$ the sufficient parameters to locate the involved wavevectors.
        Accordingly, in Eq.~\eqref{eqn:axikk}, $\hat a^\dagger_{\mkx,\mky}$ is the shorthand notation for
        \begin{equation}
          \hat a^\dagger_{\mathbf{\mko}}=\hat a^\dagger_{\mkx,\mky,\sqrt{\mko^2-\mkx^2-\mky^2}}:=\hat a^\dagger_{\mkx,\mky}\, .
          \label{eqn:adxi:sm-no(-z)}
          \end{equation}
        The two-dimensional inverse Fourier transform of~$\wpfi (k_x,k_y)$ denotes the photon-wavepacket's spatial amplitude:        
        \begin{equation}
          \begin{split}
            \wpf (x,y)& =\frac{1}{2 \pi} \iint dk_x dk_y \,  \wpfi (k_x,k_y) e^{i\left(k_x x +k_y y \right)} \, ,
          \end{split}
          \label{eqn:iftxi}
        \end{equation}
        which is the \textit{photon-wavefront}.
        Sometimes we use the term photon-wavefront instead of the equivalent term photon-wavepacket to emphasize the photon-wavepacket’s \textit{spatial representation}. 
        The corresponding two-dimensional Fourier transform of the wavefront reads        
        \begin{equation}
          \begin{split}
            \wpfi (k_x,k_y)& =\frac{1}{2 \pi} \iint dx dy \,  \wpf (x,y) e^{-i\left(k_x x +k_y y\right) } \, .
          \end{split}
          \label{eqn:ftxi}
        \end{equation}
        This paper adds the tilde sign and removes the tilde sign over a function to denote the Fourier transform and the inverse Fourier transform of the function in the spatial domain, respectively.
        \par
        Plugging Eq.~\eqref{eqn:ftxi} into the photon-wavepacket creation operator~\eqref{eqn:axikk} gives           
        \begin{equation}
            \hat a^ \dagger _{\wpf} =\iint dx dy   \,  \wpf (x,y) \hat a^\dagger_{x,y}\, ,
          \label{eqn:axixyapp}
        \end{equation}
        where                
        \begin{equation}
          \begin{split}
             \hat a^\dagger_{x,y}& =\frac{1}{2 \pi} \iint dk_x dk_y \,   \hat a^\dagger_{k_x,k_y} e^{-i\left(k_x x +k_y y \right)} \, .
          \end{split}
          \label{eqn:iftadkxy}
        \end{equation}        
        and its inverse Fourier transform is:                       
        \begin{equation}
          \begin{split}
             \hat a^\dagger_{k_x,k_y}& =\frac{1}{2 \pi} \iint dx dy \,   \hat a^\dagger_{x,y} e^{i\left(k_x x +k_y y\right) } \, .
          \end{split}
          \label{eqn:ftadxy}
        \end{equation}
        It is worth noting that since the operations and measurements usually perform in the spatial domain, we choose the wavepackets' representation in the spatial space, known as wavefront, for the subscript of the photon-wavepacket creation operators, rather than its representation in the wavevector space.
        In the above equations, $\wpf$ and $\wpfi$ represent the same photon-wavepacket in the spatial and wavevector spaces, respectively.
        However, photon-wavepacket creation operator~$\hat a^ \dagger _{\wpf}$ takes $\wpf$, the wavefront, for its subscript but not $\wpfi$ (compare Eq.~\eqref{eqn:axikk} with Eq.~\eqref{eqn:axixyapp}).
        \subsection{Multi Frequency Mode}
        \label{sec:mfm}
        The first line of Eq.~\eqref{eqn:axikk} denotes the general photon-wavepacket creation operator, where photon-wavepacket $\wpfi(\mathbf{K})$ gives the probability amplitude that the associated single-photon be in wavevector~$\mathbf{K}$.
        Above, due to the single-frequency-mode assumption and no-propagation in ($-z$)-direction assumption, the two wavevector's components~$k_x$ and~$k_y$ became sufficient parameters to locate the wavevector~$\mathbf{K}=(\mkx,\mky,\mkz)=(\mkx,\mky,\sqrt{\mko^2-\mkx^2-\mky^2})$ (see Eq.~\eqref{eqn:adxi:sm-no(-z)}).
        As a result, the volume integration was reduced to the integration over the surface, which is the upper hemisphere in Fig.~\ref{fig:sfa}.
        \par
        This section allows a single-photon to occupy multiple frequencies (wavenumbers)~$K=\frac{\omega}{c}$.
        Furthermore, it holds the no-propagation in ($-z$)-direction assumption.
        Therefore, wavenumber(wave-frequency)~$K=\frac{\omega}{c}$ together with wavevector's components~$k_x$ and~$k_y$ can locate the wavevector~$\mathbf{K}$.
        For the sake of practicality, we choose frequency~$\omega$ rather than wavenumber~$K$ to locate the wavevector~$\mathbf{K}$.
        This choice alters the photon-wavepacket creation operator~\eqref{eqn:axikk} as follows 
        \begin{equation}
          \begin{split}
            \hat a^ \dagger _{\wpf}
            &=\int  d^3K \,  \wpfi  (\mathbf{K}) \hat a ^\dagger_{\mathbf{K}}\\
            &\equiv \iiint d\omega d\mkx d\mky  \,  \wpfi (\omega,\mkx,\mky) \hat a^\dagger_{\mkx,\mky}(\omega)\, ,
            \end{split}
          \label{eqn:axiokk}
        \end{equation}
        Similar to Eq.~\eqref{eqn:iftxi} and~\eqref{eqn:ftxi}, we introduce wavefront function~$\wpf$ as the inverse Fourier transform of~$\wpfi$
        \begin{equation}
          \begin{split}           
            \wpf (\omega,x,y)& =\frac{1}{2 \pi} \iint dk_x dk_y \,  \wpfi (\omega, k_x,k_y) e^{i\left(k_x x +k_y y \right)}\\
            \wpfi (\omega,k_x,k_y)& =\frac{1}{2 \pi} \iint dx dy \,  \wpf (\omega, x,y) e^{-i\left(k_x x +k_y y\right) }   \, .
          \end{split}
          \label{eqn:ftxio}
        \end{equation}
        This paper uses the frequency-dependent photon-wavefront function ($\wpf(\omega,x,y)$) in the subscript of the photon-wavepacket creation operator~$\hat a^ \dagger _{\wpf}$ 
        \begin{equation}
          \begin{split}
           \hat a^ \dagger _{\wpf} &=\iiint d\omega dx dy   \,  \wpf (\omega,x,y) \hat a^\dagger_{x,y}(\omega)\\
           &=\iiint d\omega dk_x dk_y   \,  \wpfi (\omega,k_x,k_y) \hat a^\dagger_{k_x,k_y}(\omega)\, .
           \end{split}
          \label{eqn:axioxyapp}
        \end{equation}
        The above equation is equivalent to Eq.~\eqref{eqn:axiokk} in the wavevector space.
        \par
        For the sake of completeness, we consider the Fourier transform of the photon-wavepacket with respect to angular frequency~$\omega$, which gives the representation of the photon-wavepacket in the time domain~\cite{hayat_oe_2012}
        \small
        \begin{equation}
          \begin{split}
            \wpft (t,x,y)& =\frac{1}{\sqrt{2 \pi}} \int d\omega \,  \wpf (\omega, x,y) e^{-i\omega  t }\\
             & =\frac{1}{(2 \pi)^{\frac{3}{2}}} \iiint dk_x dk_y d\omega \wpfi (\omega, k_x,k_y) e^{i\left(k_x x +k_y y -\omega  t\right)} \\                     
             \wpfit (t,k_x,k_y)& =\frac{1}{\sqrt{2 \pi}} \int d\omega \,  \wpfi (\omega, k_x,k_y) e^{-i\omega  t }\\
            & =\frac{1}{(2 \pi)^{\frac{3}{2}}} \iiint dx dy d\omega \,  \wpf (\omega, x,y)  e^{-i\left(k_x x +k_y y +\omega  t\right) }
          \end{split}
          \label{eqn:3dftxi}
        \end{equation}
        \normalsize
        where Eq.~\eqref{eqn:ftxio} is used, and the dot sign over the function denotes the representation of the photon-wavepacket in the time domain.       
        This paper adds the dot sign and removes the dot sign over a photon-wavepacket function to represent the Fourier transform and the inverse Fourier transform of the function in the frequency domain, respectively.
        Therefore, we use four representations for the photon-wavepacket
        $\wpf (\omega, x,y),\, \wpft (t,x,y),\,  \wpfi (\omega, k_x,k_y)$ and $ \wpfit (t,k_x,k_y)$, related to each other as shown in Equations~\eqref{eqn:iftxi},\eqref{eqn:ftxi}, \eqref{eqn:ftxio} and \eqref{eqn:3dftxi}.
         Let us highlight that as Eq.~\eqref{eqn:[axi,adxi]} indicates, all the representations of the photon-wavepacket are normalized, for example
        \begin{equation}
          \begin{split}
            \iiint  d\omega dx dy  |\wpf (\omega, x,y)|^2&=\|\wpf\|^2=1\\
            \iiint  dt dk_x dk_y  |\wpfit (t,k_x,k_y)|^2 &=\|\wpfit\|^2=1\, .
            \end{split}
          \label{eqn:achi^2examp}
        \end{equation}       
        \par
        In some situations, quantum light’s spatial occupation mode becomes unimportant, such as single-mode fibers.
        For example, suppose a photon occupying a single spatial mode with central coordinate~$(\cen{x},\cen{y})$; therefore, its frequency-dependent photon-wavefront function is rewritable as $\wpf(\omega,x,y) = \wpf (\omega) \wpf_{\cen{x},\cen{y}} (x,y)$. 
        This photon-wavepacket function reduces Eq.~\eqref{eqn:axioxyapp} as follows
        \begin{equation}
          \begin{split}
            \hat a^ \dagger _{\wpf} &=\iiint d\omega dx dy   \,  \wpf (\omega)\wpf_{\cen{x},\cen{y}} (x,y) \hat a^\dagger_{x,y}(\omega)\\
            &=\int d\omega   \,  \wpf (\omega) \hat a_{\cen{x},\cen{y}}^\dagger(\omega)            \, .
            \end{split}
          \label{eqn:axioapp}
        \end{equation}
        where
        \begin{equation}
          \begin{split}
            \hat a_{\cen{x},\cen{y}}^\dagger(\omega)
            &= \iint dx dy   \, \wpf_{\cen{x},\cen{y}} (x,y)  \hat a^\dagger_{x,y}(\omega)\, .
            \end{split}
          \label{eqn:ax0y0}
        \end{equation}
        Note that the photon-wavefront of a point source at $(\cen{x},\cen{y})$ is the normalized function~$\wpf_{\cen{x},\cen{y}}(x,y)$, i.e., $\|\wpf_{\cen{x},\cen{y}}\|=1$.
        Thus, when the shape of photon-wavefront is of no account, instead of the photon-wavefront, we may write its central point at the subscript of the creation operator (compare Eq.~\eqref{eqn:axixyapp} with Eq.~\eqref{eqn:ax0y0}). 
        Furthermore, for systems that even the central point is trivial, such as  collimated quantum light or quantum light in optical fibers, the subscripts~$\cen{x}, \cen{y}$ become redundant.
        Then for simplicity, one can rewrite Eq.~\eqref{eqn:axioapp} as follows
        \begin{equation}
          \begin{split}
            \hat a^ \dagger _{\wpf}=
            & \int d\omega   \,  \wpf (\omega) \hat a^\dagger(\omega)
            \, ,
            \end{split}
          \label{eqn:1dxio}
        \end{equation}
        which is the notation used in reference~\cite{loudon_2000}.
        \par
        To sum up, suppose the photon-wavepacket function is separable into \textit{normalized}~functions of independent variables, for example, $\wpf(\omega,x,y)=\wpf(\omega) \wpf(x,y)$ with Fourier dual~$\wpfit(t,k_x,k_y)=\wpft(t) \wpfi(k_x,k_y)$.
        And imagine the considered operation depends on independent variables, for example, time evolution operator~\eqref{eqn:Utgen}, which depends only on frequency~$\omega$.
        If this is the case, one can drop the unintended variables such as in Eq.~\eqref{eqn:1dxio} and grating operation Fig.~\ref{fig:grating}.
        As another example, when the considered quantum operations are independent of frequency~$\omega$ and the photon-wavepacket is separable, we write
        \begin{equation}
          \begin{split}
            \hat a^ \dagger _{\wpf} &=\iiint d\omega dx dy   \,  \wpf (\omega,x,y) \hat a^\dagger_{x,y}(\omega)\\
            &=\iint  dx dy  \wpf(x,y) \,  \int d\omega   \wpf(\omega) \hat a^\dagger_{x,y}(\omega)\\
            &=\iint  dx dy  \wpf(x,y) \hat a^\dagger_{x,y}\, ,
            \end{split}
          \label{eqn:}
        \end{equation}
        where
        \begin{equation}
          \hat a^\dagger_{x,y}=\int d\omega   \wpf(\omega) \hat a^\dagger_{x,y}(\omega)\, .
        \end{equation}
        Such systems are the main focus of this paper, where we drop frequency~$\omega$ from consideration.
        \section{Rescaling Proportionality Relation for the Field Operators}
        The commutation relations for the spatial and wavevector field operators are as follows:
         \begin{subequations}
          \begin{align}
            \left[\hat a_{k_x,k_y},\hat a^\dagger_{k'_x,k'_y}\right]& =\delta(k_x-k'_x)\delta(k_y-k'_y) \,
            \label{eqn:[a,ad]k}\\            
            \left[\hat a_{x,y},\hat a^\dagger_{x',y'}\right]& =\delta(x-x')\delta(y-y') \, ,
          \label{eqn:[a,ad]x}
          \end{align}
          \label{eqn:[a,ad]}
        \end{subequations}
        where $\delta$ denotes the Dirac delta function.
        To obtain the rescaling proportionality relation for the field operators, we assume parameters~$u$ and~$v$ are respectively related to parameters~$x$ and $y$ with rescaling proportionality constants~$\alpha$ and $\beta$, i.e., $x=\alpha u$ and $y= \beta v$.
        Substituting these relations into Eq.~\eqref{eqn:[a,ad]x} gives
        \begin{equation}
          \begin{split}            
            \left[\hat a_{\alpha u, \beta v},\hat a^\dagger_{\alpha u', \beta v'}\right]& =\delta\left(\alpha (u-u')\right)\delta\left((\beta (v-v')\right)\\
            & =\frac{\delta\left(u-u'\right)\delta\left(v-v'\right)}{\lvert \alpha\beta \rvert }\\
            & =\frac{\left[\hat a_{u,v},\hat a^\dagger_{u',v'}\right]}{\lvert \alpha\beta \rvert}          
          \end{split}
          \label{eqn:propforad}
        \end{equation}
        where the Dirac delta function's proportionality relation~$\delta\left(\alpha (u-u')\right) =\frac{\delta\left(u-u'\right)}{\lvert \alpha \rvert} $ and Eq.~\eqref{eqn:[a,ad]} 
        are used.
        Equation~\eqref{eqn:propforad} is rewritable as $\left[\sqrt{\lvert \alpha \beta \rvert }\hat a_{\alpha u, \beta v},\sqrt{\lvert \alpha \beta \rvert  }\hat a^\dagger_{\alpha u', \beta v'}\right]=\left[\hat a_{u,v},\hat a^\dagger_{u',v'}\right] $, which denotes the rescaling proportionality relation for the field operator is
        \begin{equation}
          \hat a^{\dagger}_{\alpha x, \beta y}=\frac{1}{\sqrt{\lvert \alpha \beta\rvert }}\hat a^{\dagger}_{x,y}
          \label{eqn:propeqn}
        \end{equation}
        The field's rescaling~\eqref{eqn:propeqn} can  only be used in the context of its corresponding photon-wavepacket's \textit{integrand}.
      \section{Lens Quantum Operation}
      \label{sec:lens}
        The quantum transformation operator from the front focal plane to the back focal plane of the lens can be expressed as the products of three sequential operators, $\hat{\text{L}} =\hat{\text{T}}(f)\hat{\text{U}}_{\phi_l}  \hat{\text{T}}(f)$.
        We follow these three operations to evaluate the transformation of operator~$\hat{\text{L}}$ on photon-wavepacket creation operator~$\hat a^ \dagger _{\wpf}$.
        \par
        The first transformation, the displacement transition by distance $f$, is due to the displacement from the front focal plane up to the lens, which
        under the paraxial (Fresnel) approximation (Eq.~\eqref{eqn:tatz}) becomes
        \begin{equation}
           \begin{split}
            \hat{\text{T}}^{_{\text{Fr}}}(f) \hat a^ \dagger _{\wpf}\hat{\text{T}}^{_{\text{Fr}}\dagger}(f)
            &= \hat a^ \dagger _{\wpf^{-}}\, ,
          \end{split}
          \label{eqn:tatf}
        \end{equation}
      where the photon-wavepacket at the front of the lens in the wave-vector space is given by Eq.~\eqref{eqn:frxikk} for $z=f$: 
       \begin{equation}
          \begin{split}
            \wpfi^{-}(k_x,k_y)&=\wpfi^{_{\text{Fr}}}_{f}(k_x,k_y) \\
            &= \wpfi (k_x,k_y)  e^{i \mko f \left(1-\frac{\mkx^2}{2\mko^2}-\frac{\mky^2}{2\mko^2}\right)}\, .
          \end{split}
          \label{eqn:wfii-kxky}
          \end{equation}
          Using the Fourier transform Eq.~\eqref{eqn:iftxi}, the spatial representation (wavefront) reads:
          \begin{equation}
          \begin{split}
            \wpf^{-}(x,y)&= \frac{1}{2 \pi} \iint dk_x dk_y \,   \wpfi^{-}(k_x,k_y)\, e^{i\left(k_x x +k_y y\right) } .
            \end{split}
          \label{eqn:wfii-xy}
        \end{equation}
        \par
        The minus sign in the superscript of~$\wpf^{-}(x,y)$ indicates the photon-wavefront before entering the lens.
        After exiting the lens, the lens transforms the photon-wavepacket creation operator as follows:
        \begin{equation}
          \begin{split}
            \hat{\text{U}}_{\phi_l} \hat a^ \dagger _{\wpf^{-}}\hat{\text{U}}^{\dagger}_{\phi_l}
            &= \iint dx dy  \, \wpf^{-}(x,y)\, \hat{\text{U}}_{\phi_l} \hat a^\dagger_{x,y} \hat{\text{U}}^{\dagger}_{\phi_l}\\
            &= \iint dx dy \,  \wpf^{-}(x,y)\,   e^{-i \phi_l(x,y)} \,  \hat a^\dagger_{x,y}\\
            &= \iint dx dy \,  \wpf^{-}(x,y)\,   e^{-i \frac{\mko}{2f} (x^2+y^2)} \,  \hat a^\dagger_{x,y}\\
            &= \iint dx dy \, \wpf^{+}(x,y) \,   \hat a^\dagger_{x,y} \\
            &=a^ \dagger _{\wpf^{+}}
          \end{split}
          \label{eqn:utatu}
        \end{equation}
        where the lens phase function equation~\eqref{eqn:phil} is used, and the photon-wavefront~$ \wpf^{+}(x,y)$ (the plus superscript denotes after exiting the lens) is defined as
        \begin{equation}
          \begin{split}
            \wpf^{+}(x,y)
            &= \wpf^{-}(x,y)\,   e^{-i \frac{\mko}{2f} (x^2+y^2)} \\
            &=\frac{1}{2 \pi} \iint dk_x dk_y    \Big[        \wpfi (k_x,k_y) e^{i \mko f \left(1-\frac{\mkx^2}{2\mko^2}-\frac{\mky^2}{2\mko^2}\right) } \\
            & \phantom{=\frac{1}{2 \pi} \iint dk_x dk_y    \Big[ } \times e^{-i \frac{\mko}{2f} (x^2+y^2) +i \left(k_x x +k_y y\right) }  \Big]\, ,
          \end{split}
        \end{equation}
        where Eq.~\eqref{eqn:wfii-kxky} and \eqref{eqn:wfii-xy} are used.
        The displacement transition operator (Eq.~\eqref{eqn:Tz}) is diagonal in the wavevector basis.
        To apply the final displacement transition from the lens up to the back focal plane, therefore, using Eq.~\eqref{eqn:ftxi}, we calculate the photon-wavepacket amplitude in the wavevector basis
        \begin{equation}
          \begin{split}
            \wpfi^{+} (k'_x,k'_y)& =\frac{1}{2 \pi} \iint dx dy \,  \wpf^{+} (x,y) e^{-i\left(k'_x x +k'_y y\right) }\\
            &= \frac{-if}{\mko}   e^{i \mko f\left(1+\frac{{k'_{x}}^{2}}{2\mko^{2}}+\frac{{k'_{y}}^{2}}{2\mko^2}\right) }
            \wpf (-\frac{f}{\mko} k'_x,-\frac{f}{\mko} k'_y) \, ,
          \end{split}
          \label{eqn:iftxi+}
        \end{equation}
        where integration
        \begin{equation}
          \begin{split}
            \mbox{Int}:&= \frac{1}{2 \pi} \iint dx dy e^{-i \frac{\mko}{2f} (x^2+y^2) }   e^{-i\left( (k'_x-k_x) x +(k'_y-k_y) y\right) }  \\            
            &=\frac{-if}{\mko}e^{i \frac{f}{2\mko} ((k'_x-k_x)^2+(k'_y-k_y)^2) }\\          \end{split}
          \label{eqn:}
        \end{equation}
        is used, and Eq.~\eqref{eqn:iftxi} replaces the inverse Fourier transform of~$ \wpfi (k_x,k_y)$ at $x=-\frac{f}{\mko} k'_x$ and $y=-\frac{f}{\mko} k'_y$ with $\wpf (-\frac{f}{\mko} k'_x,-\frac{f}{\mko} k'_y)$.      
        After the displacement operation~$\hat{\text{T}}^{_{\text{Fr}}}(f)$ (Eq.~\eqref{eqn:tatf}) and the lens spatial modulation~$\hat{\text{U}}_{\phi_l}$ (Eq.~\eqref{eqn:utatu}) of the lens operator~$\hat{\text{L}}=\hat{\text{T}}^{_{\text{Fr}}}(f) \hat{\text{U}}_{\phi_l}\hat{\text{T}}^{_{\text{Fr}}}(f)$, eventually, the Fresnel displacement transition~\eqref{eqn:Tzfr} from lens to the back focal plane transforms the photon-wavepacket creation operator as follows:
        \begin{equation}
           \begin{split}
             \hat{\text{L}}  \hat a^ \dagger _{\wpf} \hat{\text{L}}^{\dagger} &=           \hat{\text{T}}^{_{\text{Fr}}}(f) \hat{\text{U}}_{\phi_l}\hat{\text{T}}^{_{\text{Fr}}}(f) \hat a^ \dagger _{\wpf}\hat{\text{T}}^{_{\text{Fr}}\dagger}(f)\hat{\text{U}}^{\dagger}_{\phi_l}\hat{\text{T}}^{_{\text{Fr}}\dagger}(f)\\
             &=           \hat{\text{T}}^{_{\text{Fr}}}(f) \hat{\text{U}}_{\phi_l} \hat a^ \dagger _{\wpf^-} \hat{\text{U}}^{\dagger}_{\phi_l}\hat{\text{T}}^{_{\text{Fr}}\dagger}(f)\\
             &=           \hat{\text{T}}^{_{\text{Fr}}}(f) \hat a^ \dagger _{\wpf^+}\hat{\text{T}}^{_{\text{Fr}}\dagger}(f)\\
             &=\iint dk'_x dk'_y \,  \wpfi^{+} (k'_x,k'_y)\, \left( \hat{\text{T}}^{_{\text{Fr}}}(f)  \hat a^\dagger_{k'_x,k'_y} \hat{\text{T}}^{_{\text{Fr}}\dagger}(f) \right) \\
             &=\iint dk'_x dk'_y \,  \wpfi^{+} (k'_x,k'_y)\,  e^{i \mko f\left(1-\frac{{\mkx'}^2}{2\mko^2}-\frac{{\mky'}^2}{2\mko^2}\right) }\,  \hat a^\dagger_{k'_x,k'_y} \\
             &= \frac{-if}{\mko}  e^{2 i \mko f} \iint dk'_x dk'_y   \wpf (-\frac{f}{\mko} k'_x,-\frac{f}{\mko} k'_y) \,  \hat a^\dagger_{k'_x,k'_y}\, .
           \end{split}
           \label{eqn:LadLk}
         \end{equation}
         where Eq.~\eqref{eqn:iftxi+} is used to substitute~$\wpfi^{+} (k'_x,k'_y)$.
         Equation~\eqref{eqn:LadLk} shows the transformation of lens operator~$\hat{\text{L}}$ on the photon-wavepacket creation operator in the wavevector space.
         Equivalently, one can write this transformation on spatial mode        
         \begin{equation}
           \begin{split}
             \hat{\text{L}}  \hat a^ \dagger _{\wpf} \hat{\text{L}}^{\dagger} 
             &=   \iint dx dy \,   \wpfi' (x, y ) \, \hat a^\dagger_{x,y} \\
             &= \hat a^ \dagger _{\wpfi'}\,             
           \end{split}
           \label{eqn:LadLx}
         \end{equation}
         where the wavefront function~$\wpfi'$ can be achieved, as Eq.~\eqref{eqn:iftxi} demonstrates, by inverse Fourier transforming the photon-wavepacket amplitude in the wavevector space of Eq.~\eqref{eqn:LadLk}, which gives
         \small
         \begin{equation}
          \begin{split}
            \wpfi' (x,y)& =\frac{1}{2 \pi} \frac{-if}{\mko}  e^{2 i \mko f}  \iint dk'_x dk'_y \,  \wpf (-\frac{f}{\mko} k'_x,-\frac{f}{\mko} k'_y)  e^{i\left(k'_x x +k'_y y \right)}\\
            & =\frac{-i\mko}{f}  e^{2 i \mko f} \left(\frac{1}{2 \pi}  \iint dk_x dk_y \,  \wpf ( k_x, k_y)  e^{-i\left(k_x \frac{\mko}{f}x +k_y \frac{\mko}{f} y \right)}\right) \\
            &= \frac{-i\mko}{f}  e^{2 i \mko f}  \wpfi ( \frac{\mko}{f}x,  \frac{\mko}{f}y )\, ,
             \end{split}
          \label{eqn:wpfi'(x,y)}
        \end{equation}
        \normalsize
        where Eq.~\eqref{eqn:ftxi} is used for the Fourier transform of wavepacket~$\wpf ( k_x, k_y)$.  
        Therefore, the lens transformation~\eqref{eqn:LadLx} indicates that the lens operator~$\hat{\text{L}}$ reshapes the photon-wavefront.
        Furthermore, as the second line of Eq.~\eqref{eqn:wpfi'(x,y)} shows, the reshaped wavefront of the lens output photon corresponds to the Fourier transform of the input wavefront.
        In other words, a lens  performs Fourier transformation on the photon-wavepacket of creation operator~$\hat a^ \dagger _{\wpf}$.
        \par 
      Due to the unitarity of the quantum Fourier optics operators~$\hat{\text{T}}^{_{\text{Fr}}}(f)\, , \ \hat{\text{U}}_{\phi_l}$ and $\hat{\text{L}}$, one can show that all photon-wavepackets~(Eq.~\eqref{eqn:wfii-kxky}, \eqref{eqn:iftxi+}, \eqref{eqn:wpfi'(x,y)}) holds the normalization condition, i.e. $\| \wpfi^{-}\|=\|\wpfi^{+}\|=\| \wpfi'\|=1$.
      \section{Quantum 4f-Processing System}
      Figure~\ref{fig:4f} shows a 4f-processing system whose operation on photon-wavepacket creation operators is expressed by Eq.~\eqref{eqn:ffadff}, $ \hat{\text{FF}}_{\phi_p}\, \hat a^ \dagger _{\wpf}\, \hat{\text{FF}}^{\dagger}_{\phi_p}=\hat a^ \dagger _{\wpfo}$.
      Let us first consider the quantum operation of the second lens of Fig.~\ref{fig:4f}, $\hat{\text{L}}  \hat a^ \dagger _{\wpfoi'} \hat{\text{L}}^{\dagger}= \hat a^ \dagger _{\wpfo}$, 
      where the relation between the output photon-wavefront~$\wpfo$ and the second lens' input photon-wavefront~$\wpfoi'$ is given by the photon-wavefront Fourier transformation~\eqref{eqn:wpfi'(x,y)}:
        \begin{equation}
          \begin{split}
            \wpfo(x,y)
            &=  \frac{-i\mko}{f}  e^{2 i \mko f}
            \frac{1}{2 \pi} \iint dk_x dk_y  \wpfoi' (k_x, k_y)  e^{-i \left(  k_x  \frac{\mko}{f}x   +  k_y  \frac{\mko}{f}y \right)}\, .
            \end{split}
          \label{eqn:wpfii'(x,y)pre}
        \end{equation}
        Also, the relation between photon-wavefront~$\wpfoi'$, the input to the second lens, and photon-wavefront~$\wpfi'(x,y)$, the output of the first lens, is given by Eq.~\eqref{eqn:wpfi''}, $ \wpfoi'(x,y)=\wpfi'(x,y) e^{-i \phi_p(x,y)}$, where $\wpfi'(x,y)$, the output of the first lens, is related to its input by the lens photon-wavefront Fourier transformation~\eqref{eqn:wpfi'(x,y)}.
        Therefore, equations~\eqref{eqn:wpfi''} and~\eqref{eqn:wpfi'(x,y)} give
        \begin{equation}
          \begin{split}
          \wpfoi'(x,y)&=\wpfi'(x,y)\, e^{-i \phi_p(x,y)} \\
          &=  \frac{-i\mko}{f}  e^{2 i \mko f}  \wpfi (\frac{\mko}{f}x, \frac{\mko}{f}y )\, e^{-i \phi_p(x,y)} \\
          &=  \frac{-i\mko}{f}  e^{2 i \mko f}  \wpfi (\frac{\mko}{f}x, \frac{\mko}{f}y )\, \Pf(x,y) \,                     
          \end{split}
          \label{eqn:}
        \end{equation}
        where $\Pf(x,y)= e^{-i \phi_p(x,y)}$ is the pupil phase factor.
        Plugging the above equation into Eq.~\eqref{eqn:wpfii'(x,y)pre} gives
        \begin{equation}
           \begin{split}
             \wpfo(x,y)
             &=  \frac{-\mko^2}{f^2}  e^{i 4 \mko f}
             \frac{1}{2 \pi} \iint dk_x dk_y \Big[ \, \wpfi ( \frac{\mko}{f} k_x,  \frac{\mko}{f} k_y) \Pf(k_x,k_y) \\
            &\phantom{=\frac{-\mko^2}{f^2}  e^{i 4 \mko f}
             \frac{1}{2 \pi} \iint}\times e^{-i \left(  k_x  \frac{\mko}{f}x  +  k_y  \frac{\mko}{f}y\right) } \Big] \\         
             &=  
             \frac{1}{2 \pi} \iint dk'_x dk'_y\, \Big[ \, \wpfi (- k'_x, - k'_y)\\
             &\phantom{=  
               \frac{1}{2 \pi} \iint }
             \times\left(-  e^{i 4 \mko f} \Pf(- \frac{f}{\mko} k'_x,- \frac{f}{\mko}k'_y) \right)  e^{i \left(  k'_x x   +  k'_y y\right) } \Big]\\
             &=\frac{1}{2 \pi} \iint dk'_x dk'_y \, \wpfi (- k'_x,  -k'_y)\, \tilde h_{\hsub}(k'_x,k'_y) \, e^{i \left(  k'_x x   +  k'_y y\right) }\\
                  &=  
              \frac{1}{2 \pi}  \iint dx' dy' \,  \wpf (-x',-y') \, h_{\hsub}(x-x',y-y')\\    
                &=  \wpf (-x,-y)\ast h_{\hsub}(x,y)\, ,
                \end{split}
          \label{eqn:4f}
        \end{equation}
      where $ k'_x= -\frac{\mko}{f} k_x$, $ k'_y=-\frac{\mko}{f} k_y$, and the transfer function of the 4f-processing system is
      \begin{equation}
        \begin{split}
          \tilde h_{\hsub}(k_x,k_y)&= -  e^{i 4 \mko f} \Pf(- \frac{f}{\mko} k_x,- \frac{f}{\mko}k_y)\\
          &= -  e^{i 4 \mko f}e^{-i \phi_p(- \frac{f}{\mko} k_x,- \frac{f}{\mko} k_y)}\, .
        \end{split}
        \label{eqn:tildh-p}
        \end{equation}
        The third line of Eq.~\eqref{eqn:4f} indicates that the Fourier transform of the 4f-operation's output photon-wavefront~$\wpfo(x,y)$ is $ \wpfi (- k_x, - k_y) \,  \tilde h_{\hsub}(k_x,k_x)$.
        Therefore, according to the convolution theorem, the 4f-operation's output photon-wavefront~$\wpfo(x,y)$ corresponds to the convolution of the inverse Fourier transform of $\wpfi ( -k_x,  -k_y)$ and $ \tilde h_{\hsub}(k_x,k_x)$, expressed in the last line of Eq.~\eqref{eqn:4f}.
        \par
        According to the scaling property of the Fourier transform, $\ft{f(ax)}=\frac{1}{\abs{a}}\tilde{f}(\frac{k_x}{a})$, the inverse Fourier transform of $\wpfi ( -k_x,  -k_y)$ is the parity transformed input photon-wavefront, $\wpf (- x, - y)$, and the inverse Fourier transform of the transfer function~$ \tilde h_{\hsub}(k_x,k_x)$ gives
        \small
        \begin{equation}
          \begin{split}
            h_{\hsub}(x,y)& =\frac{1}{2 \pi} \iint dk_x dk_y  \tilde h_{\hsub} (k_x,k_x)e^{i\left(k_x x +k_y y \right)} \\
            &= -  \frac{ e^{i 4 \mko f}}{2 \pi}  \iint dk_x dk_y \,   \Pf( -\frac{f}{\mko} k_x, -\frac{f}{\mko}k_y)  e^{i\left(k_x x +k_y y \right)} \\
            &= -  \frac{\mko^2}{f^2} \frac{ e^{i 4 \mko f}}{2 \pi}  \iint dk'_x dk'_y \,  \Pf( k'_x, k'_y)  e^{-i\left(k'_x (\frac{\mko}{f}x) +k'_y (\frac{\mko}{f} y) \right)} \\
             &= -  e^{i 4 \mko f} \frac{\mko^2}{f^2} \fPf(\frac{\mko}{f}x, \frac{\mko}{f} y) \, ,               
          \end{split}
          \label{eqn:hff}
        \end{equation}
        \normalsize
        where, $\fPf$ is the Fourier transform of the pupil phase factor~$\Pf(x,y)= e^{-i \phi_p(x,y)}$, and $h_{\hsub}$ is defined to be the impulse response (the point-spread function or the Green's function) of the 4f-processing system with pupil~$\Pf(x,y)$.
        \par
        One can show the photon-wavepacket~$\wpfo(x,y)$ at the output of the 4f-processing system for the phase-only pupil (Eq.~\eqref{eqn:4f}) is a normalized function,  $ \|\wpfo\|=1$.
        \par
        It is worth noting that, in this paper, the convolution of functions $f(t)$ and $g(t)$ is defined as follows
        \begin{equation}
          \begin{split}
            f(t)\ast g(t):&=\frac{1}{\sqrt{2\pi}} \int^{\infty}_{-\infty} f(t-\tau)g(\tau)d\tau\\
            &=\frac{1}{\sqrt{2\pi}} \int^{\infty}_{-\infty} f(\tau) g(t-\tau)d\tau\, .
          \end{split}
          \label{eqn:convdef}
          \end{equation}
      \subsection{Periodic Pupil Phase Factor}
      \label{sec:pppf}
      Assume pupil phase factor~$\Pf(x,y)=e^{-i \phi_p(x,y)}$ is a periodic function with period~$\xg$ in the x-direction and period~$\yg$ in the y-direction.
        Therefore, according to the Fourier transform theory, the phase factor's spatial frequency intervals are $\kappa_x=\frac{2 \pi }{\xg}$ and $\kappa_y=\frac{2 \pi }{\yg}$ for the $x$ and $y$ directions, respectively;
        accordingly, the phase factor is expressible as
        \begin{equation}
          \begin{split}
            \Pf(x,y) &=\sum_r\sum_s \prs_{rs} e^{i\left( r\kappa_x x + s\kappa_y y\right)} \, ,\\
          \end{split}
          \label{eqn:ft[ffp]}
        \end{equation}
        where $\prs_{rs}$ is the Fourier coefficient of pupil phase factor~$\Pf(x,y)= e^{-i \phi_p(  x, y)}$ corresponding to the two-dimensional spatial frequencies~$r \kappa_x$ and $s \kappa_y$ in $x$ and $y$ directions, respectively
        \begin{equation}
          \begin{split}
            \prs_{rs} &=\frac{1}{\xg \yg} \int^{\yg}_{0} dy \int^{\xg}_{0} dx             \Pf(x,y) e^{-i \left(r  \kappa_x x + s \kappa_y y\right)} \, .\\
          \end{split}
          \label{eqn:prs}
        \end{equation}
        \par
        Consider a 4f-processor with periodic pupil phase factor~\eqref{eqn:ft[ffp]}.
        Its impulse response function~\eqref{eqn:hff} becomes
        \begin{equation}
          \begin{split}
            h_{\hsub}(x,y)
            &= -  e^{i 4 \mko f}  \frac{1}{2 \pi}  \iint dk_x dk_y \,   \Pf( -\frac{f}{\mko} k_x, -\frac{f}{\mko}k_y)  e^{i\left(k_x x +k_y y \right)} \\
          &= -  e^{i 4 \mko f}  \frac{1}{2 \pi} \sum_{r,s} \Big[\prs_{rs}\\
          &\phantom{ - e^{i 4 \mko f}  \frac{1}{2 \pi}}\times\iint dk_x dk_y   e^{i\left(k_x ( x- r \frac{f \kappa_x}{\mko}) +k_y ( y- s  \frac{f \kappa_y}{\mko}) \right)} \Big] \\
            &= - 2 \pi  e^{i 4 \mko f}   \sum_{r,s} \prs_{rs}  \delta(x-r  \frac{f \kappa_x}{\mko}) \delta( y-s  \frac{f \kappa_y}{\mko} )  \\
            &= - 2 \pi  e^{i 4 \mko f}   \sum_{r,s} \prs_{rs}  \delta(x-r \lhx{1}) \delta( y-s \lhy{1}) \\            
            &= - 2 \pi  e^{i 4 \mko f}   \sum_{r,s} \prs_{rs}  \delta(x- \lhx{r}) \delta( y- \lhy{s})  \, ,
          \end{split}
          \label{eqn:hffdiscapp}
        \end{equation}
      where the 4f-impulse response lattice-constants $\lhx{1}$ and $\lhy{1}$ are defined as $\lhx{1}=\frac{f \kappa_x}{\mko}$
          and $\lhy{1}= \frac{f \kappa_y}{\mko}$.
      Therefore, the lattice point~$(r,s)$ has $x$-coordinate~$\lhx{r}=r\lhx{1}$ and $y$-coordinate~$\lhy{s}=s\lhy{1}$.
      Equation~\eqref{eqn:hffdiscapp} denotes a discrete power spectral density due to the periodicity assumption of the pupil phase factor.
      \par
      The 4f-processor with the lattice-like impulse response~\eqref{eqn:hffdiscapp} transforms the input photon-wavepacket~$\wpf(x,y)$ as follows (use Eq.~\eqref{eqn:4f})
      
      \begin{equation}
           \begin{split}
             \wpfo(x,y)      &=  \wpf (-x,-y)\ast h_{\hsub}(x,y)\\
             &=   \frac{1}{2 \pi}  \iint dx' dy' \,  \wpf (-x',-y') \, h_{\hsub}(x-x',y-y')          \\
             &=    -e^{i 4 \mko f}  \sum_{rs}   \prs_{rs} \iint dx' dy' \,\Big[  \wpf (-x',-y') \\
             &\phantom{=    -e^{i 4 \mko f}  \sum_{rs}   \prs_{rs} }\times
             \delta(x-x'- \lhx{r}) \delta( y-y'- \lhy{s})\Big]\\
             &=    -e^{i 4 \mko f}  \sum_{r,s}   \prs_{rs}  \wpf (-(x- \lhx{r}),-(y - \lhy{s}))  \\
             &=    -e^{i 4 \mko f}  \sum_{r,s}  \prs_{rs}  \bar{\wpf} (x- \lhx{r},y - \lhy{s})  \\
             &=    -e^{i 4 \mko f}  \sum_{r,s}  \prs_{rs}  \bar{\wpf}_{r,s}(x,y)  \, ,
                \end{split}
          \label{eqn:4fperiodic}
        \end{equation}
        where the bar sign over the wavepacket indicates the parity transformation on the wavepacket, i.e., $ \bar{\wpf} (x,y)= \wpf (-x,-y)$.
        Photon-wavepacket~$\bar{\wpf}_{r,s}=  \bar{\wpf} (x- \lhx{r},y - \lhy{s})$ has the same shape as wavepacket~$\bar{\wpf} (x,y)$, yet its origin is shifted to the lattice point~$(r,s)$ with coordinate~$(\lhx{r}, \lhy{s})$.
        Therefore, the single-photon creation operator~$\hat a^ \dagger _{\wpfo}$ with wavefront~\eqref{eqn:4fperiodic} is rewritable as the following form
          \begin{equation}
            \begin{split}
          \hat a^ \dagger _{\wpfo}
          &=\iint dx dy   \,  \wpfo (x,y) \,  \hat a^\dagger_{x,y}\\
          &=  -e^{i 4 \mko f}\sum_{r,s}  \prs_{rs}\iint dx dy   \,     \bar{\wpf}_{r,s}(x,y) \,  \hat a^\dagger_{x,y}\\
          &=  -e^{i 4 \mko f}\sum_{r,s}  \prs_{rs} \,  \hat a^\dagger_{\bar{\wpf}_{r,s}}\\          
            \end{split}
          \label{eqn:adffoutperiodic}
        \end{equation}
        \subsection{Cyclic Orthogonality of Phase Factor's Fourier Coefficients}
        \label{sec:cycortho}
        The pupil phase factor is a unit complex number at all $x$ and $y$ positions, $|\Pf(x,y)|^2= |e^{-i \phi_p(  x, y)}|^2=1$.
        This unitarity property, taking the pupil phase factor's Fourier transform~\eqref{eqn:ft[ffp]}, gives
         \begin{equation}
          \begin{split}
             |\Pf(x,y)|^2&=\sum_{\substack{r,s\\r's'}} \prs_{rs}\prs^{\ast}_{r's'} e^{i\left( r\kappa_x  x + s\kappa_y  y\right)} e^{-i\left( r'\kappa_x  x + s'\kappa_y  y\right)}\\       
            &=\sum_{\substack{r,s\\r's'}}  \prs_{rs}\prs^{\ast}_{r's'} e^{i \left(  (r-r') \kappa_x  x+  (s-s') \kappa_y  y \right)}\\
            &=\sum_{r,s}  \lvert \prs_{rs} \rvert^2
            + \sum_{\substack{r,s\\r'\neq r\\s'\neq s}} \prs_{rs}\prs^{\ast}_{r's'} e^{i \left(  (r-r') \kappa_x  x+  (s-s') \kappa_y  y \right)}\\
            &=\sum_{r,s}  \lvert \prs_{rs} \rvert^2
            + \sum_{\substack{r,s\\r''\neq 0\\s''\neq 0}} \prs_{rs}\prs^{\ast}_{r-r'',s- s''} e^{i \left(  r'' \kappa_x  x+  s'' \kappa_y  y \right)}
            \\
            &=\sum_{r,s}  \lvert \prs_{rs} \rvert^2\\
            &\phantom{=}+  \sum_{\substack{r''\neq 0\\s''\neq 0}}e^{i \left(  r'' \kappa_x  x+  s'' \kappa_y  y \right)}  \left(\sum_{r,s} \prs_{rs}\prs^{\ast}_{r-r'',s- s''}\right)\, ,
          \end{split}
        \end{equation}
        where the substitution $r''=r-r'$ and $s''=s-s'$ is used.
        Since the above equation should be one for any~$x$ and~$y$ values, the first term is one, $\sum_r\sum_s  \lvert \prs_{rs} \rvert^2=1$, and the second is zero, implying~$\sum_{r}\sum_s  \prs_{rs}\prs^{\ast}_{r-r'',s- s''}=0$.
        Therefore, we combine these two results and, in short, write 
        \begin{equation}
          \begin{split}
            \sum_{r}\sum_s  \prs_{rs}\prs^{\ast}_{r-r'',s- s''} =\delta_{r'',0}\delta_{s'',0}\, .
          \end{split}
          \label{eqn:unitaryp}
        \end{equation}
        The above equation indicates that the pupil phase factor's Fourier coefficients form a cyclic orthogonal unit vector.
        In other words, vector $\mathbf{\prs}=(\hdots,\prs_{_{-1,-1}},\prs_{_{-1,0}},\prs_{_{-1,1}}, \hdots, \prs_{_{0,-1}},\prs_{_{0,0}},\prs_{_{0,1}}, \hdots)$ is a unit vector, $|\mathbf{\prs}|^2=1$, and is orthogonal to any of its nonzero element-shifted unit vectors $\mathbf{\prs'}$, i.e., $\mathbf{\prs. \prs'}=0 $.
        For example, for an element shift of $+1$, $\mathbf{\prs'}=(\hdots,\prs_{_{-1,0}},\prs_{_{-1,1}},\prs_{_{-1,2}}, \hdots, \prs_{_{0,0}},\prs_{_{0,1}},\prs_{_{0,2}}, \hdots)$, and  $\mathbf{\prs. \prs'}=0 $.
        
        \subsection{Inverse of 4f-Processor}
        \label{app:i4f}
        Equation~\eqref{eqn:ffadff}, the unitary transformation of a 4f-processing system on a single photon creation operator, and its inverse are
        \begin{equation}
          \begin{split}
            \hat{\text{FF}}_{\phi_p}\, \hat a^ \dagger _{\wpf}\, \hat{\text{FF}}^{\dagger}_{\phi_p}&=\hat a^ \dagger _{\wpfo}\, ,\\
            \hat{\text{FF}}^{\dagger}_{\phi_p}\, \hat a^ \dagger _{\wpfo}\, \hat{\text{FF}}_{\phi_p}&=\hat a^ \dagger _{\wpf}\, .
          \end{split}
          \label{eqn:ffandffd}
        \end{equation}
        Considering Eq.~\eqref{eqn:Usi} and \eqref{eqn:ffandffd}, one can show that the quantum 4f-operator~$\hat{\text{FF}}_{\phi_p}$ transforms the quantum light~$\lvert \psi \rangle= f(\hat a_{\wpf}^ \dagger) \lvert 0 \rangle$ to quantum state~$ \hat{\text{FF}}_{\phi_p} \lvert \psi \rangle= f(\hat a_{\wpfo}^ \dagger) \lvert 0 \rangle=\lvert \Psi \rangle$, (as denoted in Eq.~\eqref{eqn:ffpsi}).
        Also, its conjugated 4f-operator~$\hat{\text{FF}}^{\dagger}_{\phi_p}$ reverses the operation, i.e., ~$ \hat{\text{FF}}^{\dagger}_{\phi_p} \lvert \Psi \rangle=\lvert \psi \rangle$.
        This section finds the relation between the pupil phase factor~$\Pf(x,y)$ of the 4f-processor and its inverse operation's pupil phase factor~$\iPf(x,y)$.
        Assume the corresponding impulse response of operator~$\hat{\text{FF}}_{\phi_p}$ ($\hat{\text{FF}}^{\dagger}_{\phi_p}$) is $h(x, y)$ ($\hi(x,y)$), which, according to Eq.~\eqref{eqn:4f}, transforms the input photon-wavepackets~$\wpfo(x,y)$ ($ \wpf(x,y)$) as follows
        \begin{equation}
          \begin{split}
            \wpfo(x,y) &=  \wpf (-x,-y)\ast h_{\hsub}(x,y)\, ,\\
            \wpf(x,y) &=  \wpfo (-x,-y)\ast \hi_{\hsub}(x,y)\, .
          \end{split}
          \label{eqn:ffffdwp}
        \end{equation}
        According to the convolution theorem, the Fourier transforms of Eq.~\eqref{eqn:ffffdwp} is
        \begin{equation}
          \begin{split}            
            \wpfoi(k_x,k_y) &=  \wpfi (-k_x,-k_y) \tilde{h}_{\hsub}(k_x,k_y)\, ,\\
            \wpfi(k_x,k_y) &=  \wpfoi (-k_x,-k_y) \thi_{\hsub}(k_x,k_y)\, ,
          \end{split}
          \label{eqn:ffffdwpft}
        \end{equation}
        which gives the following relation between the  transfer function~$ \tilde h_{\hsub}(k_x,k_x)$ of operator~$\hat{\text{FF}}_{\phi_p}$ and the transfer function~$ \thi_{\hsub}(k_x,k_x)$ of operator~$\hat{\text{FF}}^{\dagger}_{\phi_p}$.
         \begin{equation}
          \begin{split}            
             \thi_{\hsub}(k_x,k_y)=\frac{1}{\tilde{h}_{\hsub}(-k_x,-k_y)}  \, .
          \end{split}
          \label{eqn:h-h'}
        \end{equation}
        Plugging equation~\eqref{eqn:tildh-p} (ignore its trivial traveling phase factor~$e^{i 4 \mko f}$), which relates the 4f-operator's transfer function to its pupil phase factor, into Eq.~\eqref{eqn:h-h'} gives
        \begin{equation}
          \begin{split}            
            e^{-i \phi_{\ip}(- \frac{f}{\mko} k_x,- \frac{f}{\mko} k_y)}=e^{i \phi_p( \frac{f}{\mko} k_x, \frac{f}{\mko} k_y)}\, ,
          \end{split}
          \label{eqn:phi-phi'}
        \end{equation}
        where $\Pf(x,y)=e^{-i\phi_p(x,y)}$ and $\iPf(x,y)=e^{-i\phi_{\ip}(x,y)}$ are the pupil's phase factors of operator~$\hat{\text{FF}}_{\phi_p}$ and $\hat{\text{FF}}^{\dagger}_{\phi_p}$, respectively.
        Therefore, Eq.\eqref{eqn:phi-phi'} indicates that
        \begin{equation}
          \begin{split}            
            \iPf(x,y)=\Pf^{\ast}(-x,-y)\, .
          \end{split}
          \label{eqn:p-p'}
        \end{equation}
        As denoted in Eq.~\eqref{eqn:ft[ffp]}, under the periodicity assumption of the pupil phase factors, phase factors are expressible as follows
        \begin{equation}
          \begin{split}
            \Pf(x,y) &=\sum_r\sum_s \prs_{rs} e^{i\left( r\kappa_x x + s\kappa_y y\right)} \, ,\\
            \iPf(x,y) &=\sum_r\sum_s \iprs_{rs} e^{i\left( r\kappa_x x + s\kappa_y y\right)} \, .
          \end{split}
          \label{eqn:ftp-ffp'}
        \end{equation}
        Inserting Eq.~\eqref{eqn:ftp-ffp'} into Eq.~\eqref{eqn:p-p'} gives the following relation between the Fourier coefficients~$\prs_{rs}$ of the Pupil~$\Pf(x,y)$ and $\iprs_{rs}$ of $\iPf(x,y)$
        \begin{equation}
          \begin{split}
            \iprs_{rs}=\prs^{\ast}_{rs} \, .
          \end{split}
          \label{eqn:prs-prs'}
        \end{equation}
        Therefore, noting Eq.~\eqref{eqn:unitaryp} and \eqref{eqn:prs-prs'}, the Fourier coefficients of the Pupil~$\Pf(x,y)$ and $\iPf(x,y)$, $\prs_{rs}$ and ~$\iprs_{rs}$, are cyclic orthogonal to each other.
        \section{Quantum Pulse Shaping}
        \subsection{ Deterministic Pulse Shaping Conditions}
        \label{app:psuc}
        As discussed in section~\ref{sec:ps_step1} and shown in Eq.~\eqref{eqn:ffaxioffd}, a 4f-Processor with periodic pupil phase factor~\eqref{eqn:ft[ffp]mainx} maps each frequency component of the input photons into lattice-like points with frequency-dependent lattice-constant Eq.~\eqref{eqn:lxy'}
        \begin{equation}
          \lpx{1}=\frac{f c \kappa_x}{\moo}\, .
        \end{equation}              
        Still, two lattice points of two different frequencies may overlap, which we need to prevent for a perfect and deterministic quantum pulse shaping.
        \par
        The central lattice-point~$\lpx{r}=r\lpx{1}=0$ for $r=0$ is the common lattice-point for all frequencies.
        Therefore for a complete frequency separation, we assume the central lattice-point's probability amplitude~$p_{r=0}$ vanishes ($p_0=0$); in other words, the Fourier transform of the pupil phase factor~\eqref{eqn:ft[ffp]mainx} has no DC term.
        \par               
        The output lattice-constant~$\lpx{1}$ is proportional to the inverse of angular frequency $\omega$, $\lpx{1} \propto\frac{1}{\omega}$.
        Then, the higher the frequency is, the shorter its corresponding lattice-constant~$\lpx{1}$ becomes.
        Consequently, it is possible that two non-central lattice-points of two different frequency components also overlap, which we need to avoid for achieving a perfect frequency separation and then a deterministic quantum pulse shaping operation.
        Assume $\omega_{min}$ and $\omega_{max}$ are, respectively, the minimum and maximum angular frequency components that the input photons occupy, i.e., $ \wpf (\omega)=0,\  \forall  \ \omega \notin (\omega_{min},\omega_{max})$.
        The lattice-constants for these limiting frequency components are $(\lpx{1})_{max} = \frac{ f\, c\, \kappa_x}{\omega_{min}}$
        and $(\lpx{1})_{min}= \frac{ f\, c\, \kappa_x}{\omega_{max}}$, respectively.
        The min-lattice's farthest occupied lattice-point, $(\lpx{R})_{min}=R (\lpx{1})_{min}$, should be beyond the max-lattice's second farthest occupied lattice-point, $(\lpx{R-1})_{max}=(R-1) (\lpx{1})_{max}$, that is 
        \begin{equation}
          \begin{split}
            (\lpx{R-1})_{max}&< (\lpx{R})_{min}\\
            (R-1) (\lpx{1})_{max} &< R(\lpx{1})_{min}\\
             (R-1) \lpxomin{1} &< R\lpxomax{1}\\
            (R-1)\frac{f\, c\, \kappa_x}{\omega_{min}} &<R  \frac{f\, c\, \kappa_x}{\omega_{max}}\\
            (R-1)\omega_{max}&<R\omega_{min}\,            
           \end{split}
             \label{eqn:domegaps}
        \end{equation}
        which gives the deterministic pulse shaping condition as:
        \begin{subequations}
          \begin{align}
            \label{eqn:omegaratiops}       
            \frac{\omega_{max}}{\omega_{min}}&<\frac{R}{R-1}\\
           \intertext{or equivalently as}
            \label{eqn:domegaps}
            \Delta \omega &<\frac{\omega_{max}}{R}\, ,
          \end{align}          
        \end{subequations}
        where $\Delta \omega=(\omega_{max}-\omega_{min})$ is the photon spectral bandwidth.
        Note that if $R$ equals one, equation~\eqref{eqn:domegaps} becomes $\Delta \omega <\omega_{max}$, which is satisfied for any photon-wavepacket $\wpf (\omega)$.
        \par
        In conclusion, if $p_{(r=0)}=0$ and the photon bandwidth~$\Delta \omega$ is shorter than $\frac{\omega_{max}}{R}$, the 4f-processor spatially separates different spectral components of the input point source as Eq.~\eqref{eqn:ffaxioffd}.
        Therefore, it is possible to find the frequency mapped function~$\Omega(x)$, which determines the frequency component going to position~$x$.
        \subsection{Frequency Mapped Function~$\Omega(x)$}
        \label{sec:fmf}
        The $x$-coordinate of $r$th order of diffraction for frequency~$\omega$ is as follows 
        \begin{equation}
          \begin{split}
          \lpx{r} = r\lpx{1}= r\frac{f \kappa_x}{\mko}=r\frac{fc \kappa_x}{\omega}\, ,
        \end{split}
        \label{eqn:xr(omega)}
      \end{equation}
      where Eq.~\eqref{eqn:lxy'} is used.
      The $r$th order of diffraction happens with  probability amplitude~$p_{r}$ as Eq.~\eqref{eqn:ffaxioffd} indicates.
      Assume the conditions of no frequency overlap holds, meaning $r\neq0$, and $\Delta \omega <\frac{\omega_{max}}{R}$ as presented in Eq.~\eqref{eqn:domegaps}.
      Therefore, one can find function~$\Omega(x)$, which gives the mapped angular frequency at each position~$x$; therefore, $\Omega( \lpx{r})=\omega$ for $1\leq\abs{r}\leq R$.            
      To find function~$\Omega(x)$, we use Eq.~\eqref{eqn:xr(omega)}, which gives
        \begin{align}        
            &\frac{\lpx{r}}{\lpxomax{r}} =\frac{\omega_{max}}{\omega}\, , \nonumber  \\      
           \intertext{then}
            \omega &=\omega_{max}\frac{\lpxomax{r}}{\lpx{r}}\nonumber \\
            &=\omega_{max}\frac{r\lpxomax{1}}{\lpx{r}}\, .
              \label{eqn:xr/x}
        \end{align} 
      Thus, we can define the frequency mapped function~$\Omega(x)$ as follows
      \begin{equation}
          \begin{split}
            \Omega(x)
            &=\omega_{max}\frac{r(x) \lpxomax{1}}{x} \, ,
            \end{split}
        \label{eqn:Omega(x)0}
      \end{equation}
      where $r(x)$ determines which diffraction order is mapped to position~$x$.
      The smallest lattice-constant corresponds to the highest frequency ($\omega_{max}$) denoted as~$\lpxomax{1}$, which determines the cutting edge of the lowest allowed frequency.
      Furthermore, since the maximum diffraction order of the grating is~$R$, we can define diffraction order mapped function~$r(x)$ as follows 
            \begin{equation}
          \begin{split}
            & \hspace{3em}r(x)=\sgn(x) \min(R,\floor{\abs{\frac{x}{\lpxomax{1}}}})\, ,
            \end{split}
             \label{eqn:r(x)}
           \end{equation}           
           where $\sgn(x)$, $\floor{x}$, and $\abs{x}$ indicate the sign function of $x$, the floor function of $x$, and the absolute value function of $x$, respectively.
           Substitution of Eq.~\eqref{eqn:r(x)} in Eq.~\eqref{eqn:Omega(x)0} gives
         \begin{equation}
           \begin{split}
             \Omega(x)
            &=  \min(R,\floor{\frac{\abs{x}}{\lpxomax{1}}})\frac{\lpxomax{1}}{\abs{x}} \omega_{max}\, .
        \end{split}
        \label{eqn:Omega(x)app}
      \end{equation}
      Let us consider angular frequency~$\omega_0, \ \omega_{min}<\omega_0<\omega_{max}$.
      The $r$th order of diffraction maps this frequency to position~$x=\lpxoo{r}=r\lpxoo{1}	=r\frac{\lpxomax{1}}{\omega_0/\omega_{max}}$, where Eq.~\eqref{eqn:xr/x} is used.
      To inspect if $\Omega(\lpxoo{r})=\omega_0$, we plug $x=\lpxoo{r}$ in Eq.~\eqref{eqn:Omega(x)app}, which gives
      \begin{equation}
          \begin{split}
            \Omega(\lpxoo{r})
            &= \min(R,\floor{\frac{\abs{\lpxoo{r}}}{\lpxomax{1}}})\frac{\lpxomax{1}}{\abs{\lpxoo{r}}}\omega_{max} \\
            &            
            =\min(R,\floor{\frac{\abs{ r\frac{ \lpxomax{1}}{\omega_0/ \omega_{max}}}}{\lpxomax{1}}})\frac{\lpxomax{1}}{\abs{r \frac{  \lpxomax{1}}{\omega_0/ \omega_{max}}}}\omega_{max}
            \\
            &=\omega_0 \min(R,\floor{\abs{r} \frac{\omega_{max}}{\omega_0}})\frac{1}{\abs{r} }\, .
          \end{split}
          \label{eqn:Omega(x(omega0))0}
        \end{equation}
      To evaluate $\min(R,\floor{\abs{r}  \frac{\omega_{max}}{\omega_0}} )$, note that $\omega_{min}<\omega_0$, and use Eq.\eqref{eqn:omegaratiops}, which give
        \begin{equation}
          \begin{split}
            &   \abs{r}\frac{\omega_{max}}{\omega_0}<\abs{r}\frac{\omega_{max}}{\omega_{min}}<\abs{r}\frac{R}{R-1}=\abs{r}+\frac{\abs{r}}{R-1}\, .
           \end{split}
          \label{eqn:upperrangeps}
        \end{equation}
        Also, since  $\omega_0<\omega_{max}$, therefore, $\abs{r}<\abs{r}\frac{\omega_{max}}{\omega_0}$, and together with Eq.~\eqref{eqn:upperrangeps} we can write  
        \begin{equation}
          \begin{split}
               \abs{r}<\abs{r}\frac{\omega_{max}}{\omega_0}<\abs{r}+\frac{\abs{r}}{R-1}\, .
           \end{split}
          \label{eqn:domainps}
        \end{equation}
        From Eq.~\eqref{eqn:domainps}, if $\abs{r}\leq R-1$, we get $\floor{\abs{r}  \frac{\omega_{max}}{\omega_0}}=\abs{r}$, hence, $\min(R,\floor{\abs{r} \frac{\omega_{max}}{\omega_0}})=\abs{r}$.
        Also, if diffraction order~$r$ gets its extreme values~$\pm R$, i.e., $\abs{r}=R$, From Eq.~\eqref{eqn:domainps} we have $R<\abs{R}  \frac{\omega_{max}}{\omega_0}$; therefore $\min(R,\floor{\abs{R} \frac{\omega_{max}}{\omega_0}})=R=\abs{r}$.
        In conclusion, for all possible diffraction orders of the grating, i.e., $\abs{r}\leq R$, we get the following simplification 
        \begin{equation}
          \min(R,\floor{\abs{r} \frac{\omega_{max}}{\omega_0}})\frac{1}{\abs{r} }=\abs{r}\, ,
          \end{equation}
        which reduces  Eq.~\eqref{eqn:Omega(x(omega0))0} as
        \begin{equation}
          \begin{split}
            \Omega(\lpxoo{r})           
            &=\omega_0 \, ,
          \end{split}
          \label{eqn:Omega(x(omega0))}
        \end{equation}
        the proof of concept.         
\section{Simulation}
This paper presents several simulation results to illustrate the evolution of photon-wavepacket through the quantum Fourier optical systems.  
For these simulations, we consider a single photon in a Gaussian wavefront (state) or a superposition of several Gaussian wavefronts as the input of the Fourier optical system.
A python~3 script simulates the evolution of the photon-wavepacket and creates the propagating photon-wavepacket data in OpenVDB format.
The OpenVDB data file is, for higher quality and a 3d-representation of data, imported into the blender~3 software and visualized.  


\vspace{3em}
\InsertBoxL{0}{\includegraphics[width=1in,height=1.25in,clip,keepaspectratio]{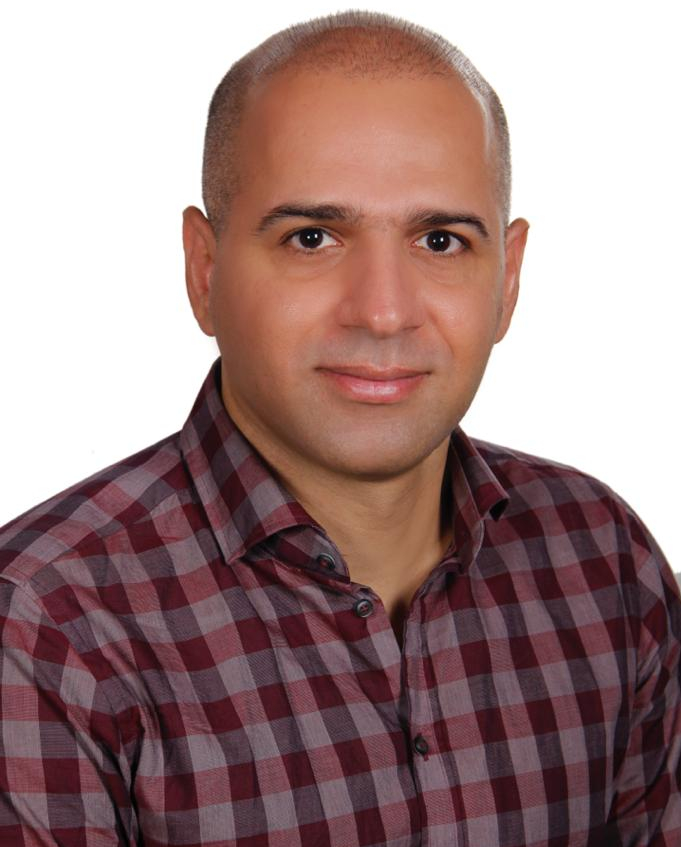}}[0]
{\footnotesize
\textbf{Mohammad Rezai} 
was born in Firoozabad, Iran in 1983.
He received the B.S. degree from the University of Sistan and Baluchestan in 2006, the M.S. degree in physics from the Sharif University of Technology in 2009, and the Ph.D. degree in physics from the University of Stuttgart, Germany, in 2018. 
From 2010 to 2013, he was a member of the International Max Planck Research School for Advanced Materials and a member of research staff in the field of condensed matter physics in the Institute for Theoretical Physics III, University of Stuttgart, Germany.
In 2013 he joined the 3rd Physikalisches Institut, University of Stuttgart, Germany, where he engaged in optical quantum information processing experiments.
\par
Since 2019, he has been a postdoctoral researcher with Sharif Quantum Center and Electrical Engineering Department, Sharif University of Technology, Tehran, Iran.
His current research interests include quantum holography, quantum Fourier optics, quantum multiple access communication systems and quantum coherence in photosynthetic systems.
\par
Dr. Rezai was elected to the Iran National Elite Foundation in 2019 and a recipient of the Max Planck scholarship in 2010. }

\vspace{3em}
\InsertBoxL{0}{\includegraphics[width=1in,height=1.25in,clip,keepaspectratio]{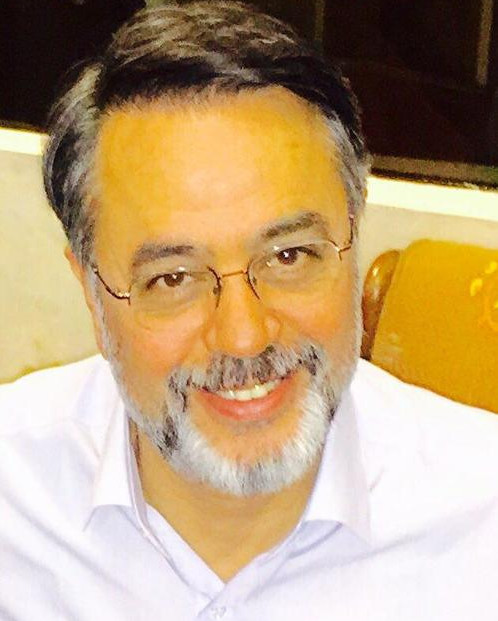}}[0]
{\footnotesize
\textbf{Jawad A. Salehi} (M’84–SM’07–F’11) was born in Kazemain, Iraq, in 1956. He received the B.Sc.degree from the University of California at Irvine in 1979, and the M.Sc. and Ph.D. degrees in electrical engineering from the University of Southern California (USC), in 1980 and 1984, respectively.
From 1984 to 1993, he was a Member of the Technical Staff of the Applied Research Area, Bell Communications Research (Bellcore), Morristown, New Jersey.
In 1990, he was with the Laboratory of Information and Decision Systems, Massachusetts Institute of Technology, as a visiting research scientist conducting research on optical multiple-access networks.
He was an Associate Professor from 1997 to 2003 and currently he is a Distinguished Professor with the Department of Electrical Engineering (EE), Sharif University of Technology (SUT), Tehran, Iran.
\par
From 2003 to 2006, he was the Director of the National Center of Excellence in Communications Science at the EE department of SUT.
In 2003, he founded and directed the Optical Networks Research Laboratory 
for advanced theoretical and experimental research in futuristic all-optical networks. 
Currently he is the Head 
of Sharif Quantum Center emphasizing in advancing quantum communication systems, quantum optical signal processing and quantum information science.
\par
His current research interests include quantum optics, quantum communications signals and systems, quantum CDMA, quantum Fourier optics, and optical wireless communication (indoors and underwater).
He is the holder of 12 U.S. patents on optical CDMA. 
\par
Dr. Salehi was named as among the 250 preeminent and most influential researchers worldwide by the Institute for Scientific Information Highly Cited in the Computer-Science Category, 2003.
He is a recipient of the Bellcore’s Award of Excellence, the Outstanding Research Award of the EE Department of SUT in 2002 and 2003, 
the Nationwide Outstanding Research Award 
2003, and the Nation’s Highly Cited Researcher Award 2004.
\par
From 2001 to 2012, he was an Associate Editor of the Optical CDMA of the IEEE TRANSACTIONS ON COMMUNICATIONS.
Professor Salehi is a member of the Iran Academy of Science and a Fellow of the Islamic World Academy of Science, Amman, Jordan.
}

\end{document}